\documentclass[printer]{aa}
\usepackage{graphicx}
\usepackage{txfonts}
\def\lesssim{\mathrel{\hbox{\rlap{\hbox{\lower4pt\hbox{$\sim$}}}\hbox{$<$}}}}
\def\gtrsim{\mathrel{\hbox{\rlap{\hbox{\lower4pt\hbox{$\sim$}}}\hbox{$>$}}}}
\newcommand{\mincir}{\raise -2.truept\hbox{\rlap{\hbox{$\sim$}}\raise5.truept
\hbox{$<$}\ }}
\newcommand{\magcir}{\raise -2.truept\hbox{\rlap{\hbox{$\sim$}}\raise5.truept
\hbox{$>$}\ }}
\newcommand{\siml}{\raise -2.truept\hbox{\rlap{\hbox{$\sim$}}\raise5.truept
\hbox{$<$}\ }}
\newcommand{\simg}{\raise -2.truept\hbox{\rlap{\hbox{$\sim$}}\raise5.truept
\hbox{$>$}\ }}
\newcommand{\be}{\begin{equation}}
\newcommand{\ee}{\end{equation}}
\newcommand{\ba}{\begin{eqnarray}}
\newcommand{\ea}{\end{eqnarray}}
\newcommand {\kpc} {$h_{70}^{-1}$ kpc$\;$}

\newcommand {\h} {$h_{70}^{-1}$ Mpc$\;$}
\newcommand {\hh} {$h_{70}^{-1}$ Mpc}
\newcommand {\hhh} {\;h_{70}^{-1} \mathrm{Mpc}}
\newcommand {\ks} {km~s$^{-1} \;$}
\newcommand {\kss} {km~s$^{-1}$}

\newcommand {\mqua} {$\times 10^{14}\;h_{70}^{-1}\;M_{\odot} \;$}

\newcommand {\mqui} {$\times 10^{15}\;h_{70}^{-1}\;M_{\odot} \;$}

\newcommand{\degree}{\ensuremath{\mathrm{^\circ}}}
\newcommand{\arcm}{\ensuremath{\mathrm{^\prime}\;}}
\newcommand{\arcs}{\ensuremath{\arcmm\hskip -0.1em\arcmm \;}}
\newcommand{\arcmm}{\ensuremath{\mathrm{^\prime}}}
\newcommand{\arcss}{\ensuremath{\arcmm\hskip -0.1em\arcmm}}
\newcommand{\dotarcs}{\,\rlap{\hbox{$\mathrm{^\prime\hskip-0.1em^\prime}$}}{\hbox{$.$}}\,}

\newcommand{\dotsec}{\,\rlap{\hbox{$\mathrm{^s}$}}{\hbox{$.$}}\,}
\begin{document}
   \title{Optical analysis of the poor clusters \\
Abell~610, Abell~725, and Abell~796, \\ containing diffuse radio sources}
%   \subtitle{}
%
\author{W. Boschin\inst{1,2}
\and R. Barrena\inst{3} 
\and M. Girardi\inst{2,4}
\and M. Spolaor\inst{2,5}
}
   \offprints{W. Boschin; e.mail: boschin@tng.iac.es}

\institute{
Fundaci\'on Galileo Galilei -- INAF, Rambla Jos\'e Ana Fern\'andez Perez 7, E--38712 Bre\~na Baja (San Antonio), Canary Islands, Spain\\
\and
Dipartimento di Astronomia, Universit\`{a} degli Studi di Trieste, via Tiepolo 11, I--34143 Trieste, Italy\\
\and
Instituto de Astrofisica de Canarias, C/Via Lactea s/n, E--38205 La Laguna, Tenerife, Canary Islands, Spain\\
\and 
INAF -- Osservatorio Astronomico di Trieste, via Tiepolo 11, I--34143  Trieste, Italy\\
\and
Centre for Astrophysics \& Supercomputing, Swinburne University, Hawthorn, VIC 3122, Australia\\
}

\date{Received / accepted }

\abstract{ }{We study the dynamical status of the poor, low X--ray
luminous galaxy clusters Abell 610, Abell 725, and Abell 796 (at
$z=0.1$, 0.09, and 0.16, respectively), containing diffuse radio
sources (relic, relic, and possible halo, respectively).}
{Our analysis is based on new spectroscopic data obtained at the
William Herschel Telescope for 158 galaxies, new photometry obtained
at the Isaac Newton Telescope with the addition of data recovered from
the Data Release 5 of the Sloan Digital Sky Survey.  We use
statistical tools to select 57, 36, and 26 cluster members and to
analyze the kinematics of cluster galaxies, as well as to study the 2D
cluster structure.}{The low values we compute for the global
line--of--sight velocity dispersion of galaxies ($\sigma_{\rm
V}=420-700$ \kss) confirm that these clusters are low--mass clusters.
Abell 610 shows a lot of evidence of substructure.  It seems to be
formed by two structures separated by $\sim 700$ \ks in the cluster
rest--frame, having comparable $\sigma_{\rm V} \sim 200$ \ks and
likely causing a velocity gradient. The velocity of the brightest
cluster member (BCMI; a bright radio source) is very close to the mean
velocity of the higher velocity structure.  A third small,
low--velocity group hosts the second brightest cluster member (BCMII).
The analysis of the 2D galaxy distribution shows a bimodal
distribution in the core elongated in the SE--NW direction and likely
associated to BCMI and BCMII groups. Abell 725 and Abell 796, which
are less sampled, show marginal evidence of substructure in the
velocity space. They are elongated in the 2D galaxy distribution.  For
both Abell 610 and Abell 725 we shortly discuss the possible
connection with the hosted diffuse radio relic.}{Our results show that
relic radio sources are likely connected with merger events,
but are not limited to massive clusters. About the possible halo
source in Abell 796, there is some evidence of a merger event in this
non--massive cluster, but a pointed radio observation is necessary to
confirm this halo.}


\authorrunning{Boschin et al.}
\titlerunning{Optical analysis of poor radio clusters} 
\maketitle
%
%________________________________________________________________

\section{Introduction}

Clusters of galaxies are not recognized as simple relaxed
structures, but rather as evolving via merging processes in a
hierarchical fashion from poor groups to rich clusters.  A recent
aspect of these investigations is the possible connection of cluster
mergers with the presence of extended, diffuse radio sources: halos
and relics. These radio sources are large (up to $\sim1$ \hh),
amorphous cluster sources of uncertain origin and generally steep
radio spectra (Hanisch \cite{han82}; see also Giovannini \& Feretti
\cite{gio02} for a more recent review).  They are rare sources that
appear to be associated with very rich clusters that have undergone
recent mergers. Therefore, it has been suggested by various authors
that cluster halos/relics are related to recent merger activity (e.g.,
Tribble \cite{tri93}; Burns et al. \cite{bur94}; Feretti
\cite{fer99}).

The synchrotron radio emission of halos and relics demonstrates the
existence of large scale cluster magnetic fields, of the order of
0.1--1 $\mu$G, and of widespread relativistic particles of energy
density 10$^{-14}$ -- 10$^{-13}$ erg cm$^{-3}$.  The difficulty in
explaining radio halos/relics arises from the combination of their
large size and the short synchrotron lifetime of relativistic
electrons.  The expected diffusion velocity of the electron population
is of the order of the Alfven speed ($\sim$100 \kss), making it
difficult for the electrons to diffuse over a megaparsec--scale region
within their radiative lifetime. Therefore, one needs a mechanism by
which the relativistic electron population can be transported over
large distances in a short time, or a mechanism by which the local
electron population is reaccelerated and the local magnetic fields are
amplified over an extended region.  The cluster--cluster merger can
potentially supply both mechanisms (e.g., Giovannini et
al. \cite{gio93}; Burns et al. \cite{bur94}; R\"ottgering et
al. \cite{rot94}; see also Feretti et al. \cite{fer02a}; Sarazin
\cite{sar02} for reviews). However, the question is still debated
since the diffuse radio sources are quite uncommon and only recently
have we been able to study these phenomena on the basis of sufficient
statistics (few dozen clusters up to $z\sim0.3$, e.g., Giovannini et
al. \cite{gio99}; see also Giovannini \& Feretti \cite{gio02}; Feretti
\cite{fer05}).

Growing evidence of the connection between diffuse radio emission and
cluster merging is based on X--ray data (e.g., B\"ohringer \& Schuecker
\cite{boh02}; Buote \cite{buo02}). Studies based on a large number of
clusters have found a significant relation between the radio and the
X--ray surface brightness (Govoni et al. \cite{gov01a}, \cite{gov01b})
and between the presence of radio halos/relics and irregular and
bimodal X--ray surface brightness distribution (Schuecker et
al. \cite{sch01}).

Optical data are a powerful way to investigate the presence and the
dynamics of cluster mergers, too (e.g., Girardi \& Biviano
\cite{gir02}). The spatial and kinematical analysis of member galaxies
allow us to detect and measure the amount of substructure, to identify
and analyze possible pre--merging clumps or merger remnants.  This
optical information is complementary to X--ray information since
galaxies and intracluster medium react on different timescales
during a merger (see, e.g., numerical simulations by Roettiger et
al. \cite{roe97}).  Unfortunately, to date optical data are lacking or
are poorly exploited to investigate the phenomenon of diffuse radio
sources.  The sparse literature contains a few individual clusters
(e.g., Barrena et al. \cite{bar02}; Mercurio et al. \cite{mer03};
Maurogordato et al. \cite{mau08}).

In this context, we are conducting an intensive observational and
data analysis program to study the internal dynamics of radio
clusters by using member galaxies. Clusters already analyzed are:
Abell 2219 (Boschin et al. \cite{bos04}); Abell 2744 (Boschin et
al. \cite{bos06}); Abell 697 (Girardi et al. \cite{gir06}); Abell 773
(Barrena et al. \cite{bar07a}); Abell 115 (Barrena et
al. \cite{bar07b}). These are all massive clusters and, indeed, to
date analyzed clusters containing a radio halo or a relic source have
a large gravitational mass (larger than $0.7$ \mqui within 2\hh; see
Giovannini \& Feretti \cite{gio02}). From the theoretical point of
view, a large mass is an expected property in radio clusters since the
energy available to accelerate relativistic particles in a merger
scales as $\sim M^2$, as discussed by Buote (\cite{buo01}).

In this paper, we report our results about three poor Abell clusters:
Abell 610, Abell 725, and Abell 796 (hereafter A610, A725, and A796)
having Abell richness (i.e., the number of galaxies within 1 Abell
radius and magnitude between $m_3$ and $m_{3}+2$; see Abell et
al. \cite{abe89}) of 46, 36, and 56 galaxies respectively. These
clusters contain a radio relic or a (possible) radio
halo. In particular, A610 exhibits a relic source confirmed by
pointed observations (Giovannini \& Feretti \cite{gio00}); A725
presents a relic source well visible in survey data (Kempner \&
Sarazin \cite{kem01}); and A796 possibly hosts a halo source visible
in survey data (Kempner \& Sarazin \cite{kem01}) which needs to be
confirmed. To date, no measure of internal velocity dispersion and/or
X--ray temperature have been reported in the literature.  We have
recently carried out spectroscopic observations with the William
Herschel Telescope giving new redshift data for 158 galaxies in the
field of these clusters, as well as photometric observations at the
Isaac Newton Telescope for A725. We recover additional photometric and
spectroscopic information from the Data Release 5 of the Sloan Digital
Sky Survey (SDSS DR5).

The paper is organized as follows. We present the new optical data in
Sect.~2. We present the relevant analyses and conclusions in Sects.~3,
4, and 5 for A610, A725, and A796, respectively.  We summarize and
discuss our results in Sect.~6.

Unless otherwise stated, we give errors at the 68\% confidence level
(hereafter c.l.).  Throughout the paper, we assume a flat cosmology
with $\Omega_{\rm m}=0.3$, $\Omega_{\Lambda}=0.7$ and $H_0=70$
$h_{70}$ \ks Mpc$^{-1}$. For this cosmological model, 1\arcm
corresponds to 108, 102, and 163 \kpc at the A610, A725, and A796
redshift.

\section{Data sample}

\subsection{Spectroscopy}

We carried out multi--object spectroscopic observations of A610, A725,
and A796 in November 2004 and January 2005. We used AF2/WYFFOS, the
multi--object, wide--field, fiber spectrograph working at the prime
focus of the 4.2 m William Herschel Telescope (WHT, Island of La
Palma, Spain). This spectrograph is equipped with optical fibers each
of 1.6 arcsec diameter. We used the grism R300B in combination with
the 2--chip EEV 4K$\times$4K pixel mosaic (pixel size 13.5 $\mu$m)
working in binning 2$\times$2. In the case of A610 and A725, we
acquired two exposures of 1800 s for two fiber configurations, in the
case of A796 we acquired three exposures of 1800 s for two fiber
configurations. We performed wavelength calibration with the helium
lamp. Reduction of spectroscopic data was carried out with the IRAF
package.
\footnote{IRAF is distributed by the National Optical Astronomy
Observatories, which are operated by the Association of Universities
for Research in Astronomy, Inc., under cooperative agreement with the
National Science Foundation.}

We determined radial velocities with the cross--correlation
technique (Tonry \& Davis \cite{ton79}) implemented in the RVSAO
package (developed at the Smithsonian Astrophysical Observatory
Telescope Data Center).  Each spectrum was correlated against six
templates for a variety of galaxy spectral types: E, S0, Sa, Sb, Sc
and Ir (Kennicutt \cite{ken92}). The template producing the highest
value of $\cal R$, i.e., the parameter given by RVSAO and related to
the signal--to--noise of the correlation peak, was chosen.  Moreover,
all the spectra and their best correlation functions were examined
visually to verify the redshift determination.  In one case (galaxy
ID~127 of A610, see Table~\ref{catalog610}), we took the EMSAO redshift
as a reliable estimate of the redshift. We obtained redshifts for 62,
51, 45 galaxies for A610, A725, A796.

For ten galaxies we obtained two redshift determinations of similar
quality.  This allows us to obtain a more rigorous estimate for the
redshift errors since the nominal errors as given by the
cross--correlation are known to be smaller than the true errors (e.g.,
Malumuth et al.  \cite{mal92}; Bardelli et al. \cite{bar94}; Ellingson
\& Yee \cite{ell94}; Quintana et al. \cite{qui00}).  For these ten
galaxies we fit the first determination vs. the second one by using a
straight line and considering errors in both coordinates (e.g., Press
et al. \cite{pre92}). The fitted line agrees with the one--to--one
relation, but, when using the nominal cross--correlation errors, the
small value of the $\chi^2$ probability indicates a poor fit,
suggesting the errors are underestimated.  Only when nominal errors
are multiplied by a factor of $\sim$1.3 can the observed scatter be
justified. We therefore assume hereafter that true errors are larger
than nominal cross--correlation errors by a factor 1.3. For the ten
galaxies we used the weighted mean of the two redshift determinations
and the corresponding error.

For A610 and A796 we add galaxies with spectroscopic and/or
photometric data found in the SDSS DR5. For A610, we find 147 galaxies
within a radius of 30\arcm from the cluster center given by Abell et
al. (\cite{abe89}).  This radius is about the double of that sampled
by our data and well larger than the virial radius of the cluster (see
Sect.~3.1), and thus will be useful to study the cluster periphery,
too.  Out of 147 SDSS galaxies, 44 are in common with our WHT
galaxies.  We fit SDSS redshift determination vs. our determination
finding that the fitted line well agrees with the one--to--one
relation. Again, the small value of the $\chi^2$ probability indicates
a poor fit, suggesting the errors are underestimated, but we prefer to
not apply any correction since the measurements come from two
different sources.  We combine WHT and SDSS data using the weighted
mean of the two redshift determinations and the corresponding error.
Our final spectroscopic catalog consists of 165 galaxies. As for A796,
we find 72 SDSS galaxies -- out of which there are 18 in common with
our WHT galaxies -- and compile a final spectroscopic catalog of 99
galaxies.

\subsection{Photometry}

Our photometric observations were carried out with the Wide Field
Camera (WFC), mounted at the prime focus of the 2.5 m Isaac Newton
Telescope (INT, Island of La Palma, Spain). We observed A725 in
December 2004 in photometric conditions with a seeing of about
1.8\arcs in $R_{\rm H}$ (Harris) band and 2.5\arcs in $B_{\rm H}$.

The WFC consists of a four--CCD mosaic covering a
33\arcmm$\times$33\arcm field of view, with only a 20\% marginally
vignetted area. We took 9 exposures of 720 s in $B_{\rm H}$ and
another 13 exposures of 360 s in $R_{\rm H}$ filters (a total of 6480
s and 4680 s in the two bands, respectively). We developed a dithering
pattern to build a ``supersky'' frame that was used to correct our
images for fringing patterns (Gullixson \cite{gul92}). In addition,
the dithering helped us to clean cosmic rays and avoid gaps between
the CCDs in the final images. The complete reduction process
(including flat fielding, bias subtraction, and bad--column
elimination) yielded a final coadded image where the variation of the
sky was lower than 1.0\% within a region of 13\arcm radius from the
center of the cluster.

In order to match the photometry of several filters, a good
astrometric solution is needed. The astrometry has to take into
account the field distortions present in the WFC full frame. Using
IRAF tasks and taking as a reference the USNO B1.0 catalog, we were
able to find an accurate astrometric solution (rms $\sim$ 0.5\arcss)
across the full frame.

We performed the photometric calibration with Landolt standard fields
observed in December 2007 using the 80 cm telescope at Teide
Observatory. We finally identified galaxies in our $B_{\rm H}$ and
$R_{\rm H}$ images and measured their magnitudes with the SExtractor
package (Bertin \& Arnouts \cite{ber96}) and the AUTOMAG procedure. In
a few cases (e.g., objects close to very bright and saturated stars)
the standard SExtractor photometric procedure failed. In these cases,
by performing an isophotal analysis (task {\it bmodel} in IRAF) we
eliminate the bright objects. Then, we reran SExtractor to estimate
the magnitudes of faint targets close to bright stars.

We transformed all Harris magnitudes into the Johnson--Cousins system
(Johnson \& Morgan \cite{joh53}; Cousins \cite{cou76}). We used
$B=B\rm_H+0.13$ and $R=R\rm_H$ as derived from the Harris filter
characterization
(http://www.ast.cam.ac.uk/$\sim$wfcsur/technical/photom/colours/) and
assuming a $B-V\sim 1.0$ for E--type galaxies (Poggianti \cite{pog97}).
As a final step, we estimated and corrected the galactic extinction
$A_B \sim0.23$, $A_R \sim0.14$ and $E(B-V)=0.05$ from Burstein \&
Heiles's (\cite{bur82}) reddening maps.

The photometric sample of galaxies in the field of A725 is complete
down to $B=22.5$ (24.4) and $R=20.6$ (22.1) for $S/N=5$ (3) within a
square region of 33\arcmm$\times$33\arcm centered on
R.A.=$09^{\mathrm{h}}00^{\mathrm{m}}50^{\mathrm{s}}$ and
Dec.=$+62\degree 39\arcmm 00\arcs$ (J2000), and with a void of
11\arcmm$\times$11\arcm in the NW of the field. As for the
spectroscopic sample, it is $\sim$25\% complete down to $R=17.5$ in
the whole field and $\sim$45\% complete down to $R=17.5$ within 1
$\mathrm{Mpc \;}$from the cluster center.

For A610 and A796, we use public photometric data from the SDSS DR5.
In particular, we use $r'$, $i'$, $z'$ magnitudes, already corrected
for the Galactic extinction and consider galaxies within a radius of
30\arcm from the cluster center given by Abell et
al. (\cite{abe89}). In the case of A610, our spectroscopic sample is
$\sim$40\% complete down to $r'=17.5$ within 1 $\mathrm{Mpc \;}$from
the cluster center. As for A796, our spectroscopic sample is
$\sim$20\% complete down to $r'=19$ within 1 $\mathrm{Mpc \;}$from
the cluster center.

\section{Abell 610}

\begin{figure}
%\centering
\resizebox{\hsize}{!}{\includegraphics{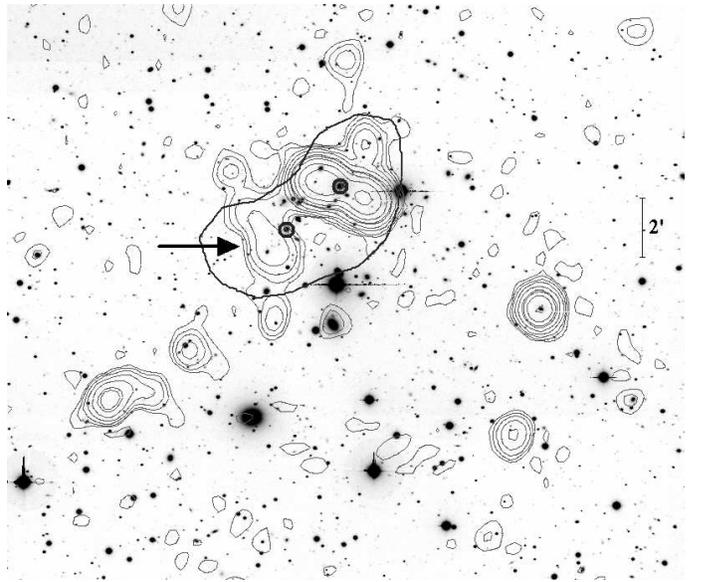}}
\caption{$R$--band image of A610 (data taken with the WFC camera of
the INT) with, superimposed, the contour levels of a VLA radio image
at 20 cm (thin contours; Giovannini \& Feretti \cite{gio00}). To avoid
confusion, only one isodensity contour (thick contour) of the spatial
distribution of the likely cluster members (see text) is shown. Two
thick circles highlight the positions of the two brightest cluster
members. The arrow shows the position of the radio relic. North is at
the top and east to the left.}
\label{A610VLA}
\end{figure}

\begin{figure*}
%\centering
\resizebox{\hsize}{!}{\includegraphics{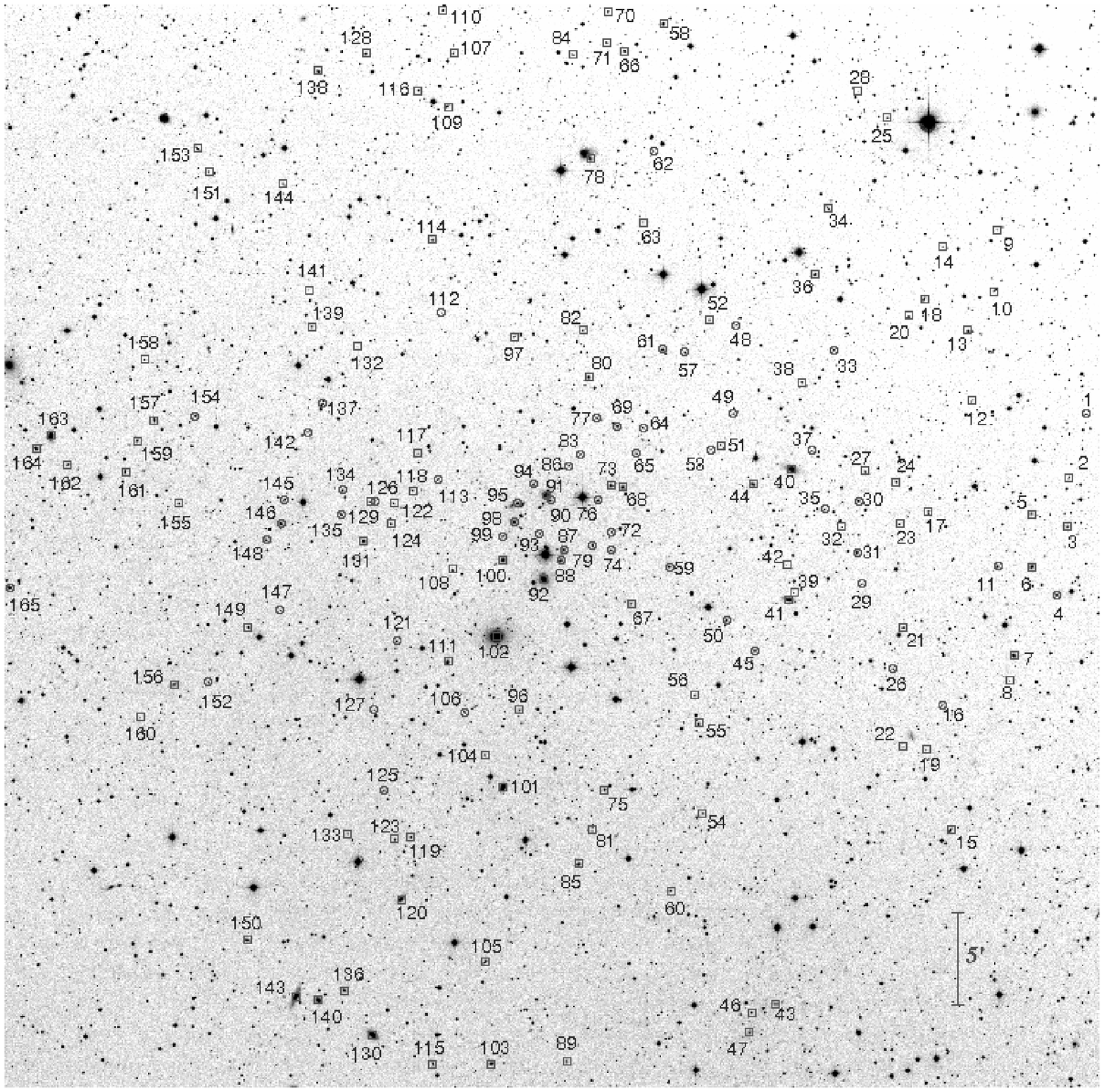}}
\caption{DSS2 $R$--band image of A610 (North at the top and East to the
left). Galaxies with successful velocity measurements are labeled as
in Table~\ref{catalog610}. Circles and boxes indicate cluster members and
non--member galaxies, respectively.}
\label{figimage610}
\end{figure*}

%\documentclass{aa}
%\usepackage{graphicx}
%%new commands
%\def\lesssim{\mathrel{\hbox{\rlap{\hbox{\lower4pt\hbox{$\sim$}}}\hbox{$<$}}}}
%\def\gtrsim{\mathrel{\hbox{\rlap{\hbox{\lower4pt\hbox{$\sim$}}}\hbox{$>$}}}}
%\newcommand{\mincir}{\raise -2.truept\hbox{\rlap{\hbox{$\sim$}}\raise5.truept
%\hbox{$<$}\ }}
%\newcommand{\magcir}{\raise -2.truept\hbox{\rlap{\hbox{$\sim$}}\raise5.truept
%\hbox{$>$}\ }}
%\newcommand{\siml}{\raise -2.truept\hbox{\rlap{\hbox{$\sim$}}\raise5.truept
%\hbox{$<$}\ }}
%\newcommand{\simg}{\raise -2.truept\hbox{\rlap{\hbox{$\sim$}}\raise5.truept
%\hbox{$>$}\ }}
%\newcommand{\be}{\begin{equation}}
%\newcommand{\ee}{\end{equation}}
%\newcommand{\ba}{\begin{eqnarray}}
%\newcommand{\ea}{\end{eqnarray}}
%\newcommand {\h} {$h^{-1}$ Mpc $ \;$}
%\newcommand {\kpc} {$h^{-1}$ kpc}
%\newcommand {\hh} {$h^{-1}$ Mpc}
%\newcommand {\ks} {km~s$^{-1} \;$}
%\newcommand {\kss} {km~s$^{-1}$}
%\newcommand {\mpc} {$Mpc \;$}
%\newcommand {\msun} {$h^{-1} \  M_{\odot} \;$}
%\newcommand {\m} {$M_{\odot} \;$}
%\newcommand {\ml} {$h \, M_{\odot}/L_{\odot} \;$}
%\newcommand {\mll} {$h \, M_{\odot}/L_{\odot}$}
%\newcommand{\vel}{\,{\rm km\,s^{-1}}}
%%
%\begin{document}
%
%\addtocounter{table}{-2}
\begin{table}[!ht]
        \caption[]{Velocity catalog of 165 spectroscopically measured
galaxies in the field of A610. In Col.~1, IDs in italics indicate
non--cluster galaxies. IDs 91 and 98 (in boldface) are, respectively, the
first and the second brightest galaxies of the cluster. Asterisks in
Col.~1 highlight QSOs in the SDSS catalog.}
         \label{catalog610}
              $$ 
        % \begin{array}{p{0.5\linewidth}l}
           \begin{array}{r c c r r c}
            \hline
            \noalign{\smallskip}
            \hline
            \noalign{\smallskip}

\mathrm{ID} & \mathrm{\alpha},\mathrm{\delta}\,(\mathrm{J}2000)  & r^\prime & \mathrm{v}\,\,\,\,\,\,\, & \mathrm{\Delta}\mathrm{v} & \mathrm{Source}\\
  & & &\mathrm{(\,km}&\mathrm{s^{-1}\,)}& \\
            \hline
            \noalign{\smallskip}  

1             & 7\ 57\ 04.15 ,+27\ 13\ 13.3&       17.65&  29052 &  16 & \mathrm{S}\\
\textit{2}    & 7\ 57\ 08.66 ,+27\ 09\ 46.4&       17.51&  19274 &  28 & \mathrm{S}\\
\textit{3}    & 7\ 57\ 09.21 ,+27\ 07\ 06.9&       16.71&  34657 &  50 & \mathrm{S}\\
4             & 7\ 57\ 12.09 ,+27\ 03\ 23.8&       16.32&  29776 &  46 & \mathrm{S}\\
\textit{5}    & 7\ 57\ 18.01 ,+27\ 07\ 51.1&       16.92&  19291 &  44 & \mathrm{S}\\
\textit{6}    & 7\ 57\ 18.16 ,+27\ 04\ 57.8&       15.72&  19913 &  45 & \mathrm{S}\\
\textit{7}    & 7\ 57\ 22.92 ,+27\ 00\ 08.5&       15.93&  19633 &  43 & \mathrm{S}\\
\textit{8}*   & 7\ 57\ 24.02 ,+26\ 58\ 46.9&       18.94& 720155 & 395 & \mathrm{S}\\
\textit{9}    & 7\ 57\ 25.29 ,+27\ 23\ 21.9&       18.37&  30870 &  40 & \mathrm{S}\\
\textit{10}   & 7\ 57\ 26.27 ,+27\ 20\ 01.3&       17.22&  37076 &  39 & \mathrm{S}\\
11            & 7\ 57\ 26.55 ,+27\ 05\ 00.3&       16.58&  29492 &  41 & \mathrm{S}\\
\textit{12}   & 7\ 57\ 32.19 ,+27\ 14\ 03.1&       17.36&  20638 &  23 & \mathrm{S}\\
\textit{13}   & 7\ 57\ 32.84 ,+27\ 17\ 51.3&       16.40&  37190 &  44 & \mathrm{S}\\
\textit{14}   & 7\ 57\ 38.68 ,+27\ 22\ 29.0&       18.42&  69538 &  43 & \mathrm{S}\\
\textit{15}   & 7\ 57\ 39.03 ,+26\ 50\ 42.6&       16.38&   8076 &  16 & \mathrm{S}\\
16            & 7\ 57\ 40.71 ,+26\ 57\ 28.9&       17.70&  29271 &  18 & \mathrm{S}\\
\textit{17}   & 7\ 57\ 43.25 ,+27\ 08\ 04.3&       17.52&  36975 &  46 & \mathrm{S}\\
\textit{18}   & 7\ 57\ 43.57 ,+27\ 19\ 38.2&       16.15&  37018 &  46 & \mathrm{S}\\
\textit{19}   & 7\ 57\ 44.76 ,+26\ 55\ 06.3&       19.09& 135715 &  64 & \mathrm{S}\\
\textit{20}   & 7\ 57\ 47.25 ,+27\ 18\ 42.7&       17.38&  42430 &  46 & \mathrm{S}\\
\textit{21}   & 7\ 57\ 49.87 ,+27\ 01\ 45.5&       16.95&  21660 &  15 & \mathrm{S}\\
\textit{22}*  & 7\ 57\ 50.40 ,+26\ 55\ 14.7&       18.69& 689658 & 435 & \mathrm{S}\\
\textit{23}   & 7\ 57\ 50.44 ,+27\ 07\ 22.9&       17.60&  21562 &  36 & \mathrm{W+S}\\
\textit{24}   & 7\ 57\ 51.13 ,+27\ 09\ 41.8&       17.71&  13929 &  12 & \mathrm{W+S}\\
\textit{25}   & 7\ 57\ 51.74 ,+27\ 29\ 33.5&       17.23&  62598 &  49 & \mathrm{S}\\
26            & 7\ 57\ 52.54 ,+26\ 59\ 31.1&       17.40&  29769 &  40 & \mathrm{W+S}\\
\textit{27}   & 7\ 57\ 58.64 ,+27\ 10\ 17.0&       17.05&  21681 &  19 & \mathrm{W+S}\\
\textit{28}*  & 7\ 57\ 58.93 ,+27\ 31\ 02.7&       19.00& 431857 & 470 & \mathrm{S}\\
29            & 7\ 57\ 59.88 ,+27\ 04\ 09.2&       17.38&  29630 &  25 & \mathrm{S}\\
30            & 7\ 58\ 00.36 ,+27\ 08\ 41.9&       16.93&  28778 &  37 & \mathrm{S}\\
31            & 7\ 58\ 00.96 ,+27\ 05\ 53.8&       16.12&  28890 &  30 & \mathrm{W+S}\\
\textit{32}*  & 7\ 58\ 04.51 ,+27\ 07\ 18.2&       19.12& 448615 & 286 & \mathrm{S}\\
33            & 7\ 58\ 05.84 ,+27\ 16\ 56.9&       16.58&  28919 &  30 & \mathrm{W+S}\\
\textit{34}   & 7\ 58\ 06.68 ,+27\ 24\ 40.2&       16.93&   7769 &  30 & \mathrm{S}\\
35            & 7\ 58\ 08.69 ,+27\ 08\ 15.5&       17.59&  28924 &  33 & \mathrm{W+S}\\

               \noalign{\smallskip}			    
            \hline					    
            \noalign{\smallskip}			    
            \hline					    
         \end{array}
     $$ 
         \end{table}
\addtocounter{table}{-1}
\begin{table}[!ht]
          \caption[ ]{Continued.}
     $$ 
           \begin{array}{r c c r r c}
            \hline
            \noalign{\smallskip}
            \hline
            \noalign{\smallskip}

\mathrm{ID} & \mathrm{\alpha},\mathrm{\delta}\,(\mathrm{J}2000)  & r^\prime & \mathrm{v}\,\,\,\,\,\,\, & \mathrm{\Delta}\mathrm{v}& \mathrm{Source} \\
%  &                      &  &  &\,\,\,\,\,\,\,\mathrm{(\,km}&\mathrm{s^{-1}\,)}\,\,\,&  & & \\
  & & &\mathrm{(\,km}&\mathrm{s^{-1}\,)}& \\
            \hline
            \noalign{\smallskip}

\textit{36}   & 7\ 58\ 10.17 ,+27\ 21\ 04.7&       16.57&   6841 &  45 & \mathrm{S}\\
37            & 7\ 58\ 11.69 ,+27\ 11\ 28.6&       17.46&  29287 &  31 & \mathrm{W+S}\\
\textit{38}   & 7\ 58\ 13.82 ,+27\ 15\ 12.7&       17.54&  23099 &  14 & \mathrm{S}\\
\textit{39}   & 7\ 58\ 16.58 ,+27\ 03\ 43.9&       18.40&  43608 & 101 & \mathrm{W}\\
\textit{40}   & 7\ 58\ 16.66 ,+27\ 10\ 29.6&       14.51&  13818 &  39 & \mathrm{W+S}\\
\textit{41}   & 7\ 58\ 17.86 ,+27\ 03\ 23.9&       14.68&   6386 &  24 & \mathrm{W+S}\\
\textit{42}*  & 7\ 58\ 18.29 ,+27\ 05\ 16.5&       19.89& 740173 & 237 & \mathrm{S}\\
\textit{43}   & 7\ 58\ 22.66 ,+26\ 41\ 20.6&       17.48&  68989 &  38 & \mathrm{S}\\
\textit{44}   & 7\ 58\ 26.22 ,+27\ 09\ 41.5&       15.74&  14116 &  17 & \mathrm{W+S}\\
45            & 7\ 58\ 26.37 ,+27\ 00\ 37.5&       16.63&  30012 &  39 & \mathrm{S}\\
\textit{46}   & 7\ 58\ 28.45 ,+26\ 40\ 53.9&       17.67&  24891 &  42 & \mathrm{S}\\
\textit{47}   & 7\ 58\ 28.97 ,+26\ 39\ 49.6&       16.77&  70367 &  52 & \mathrm{S}\\
48            & 7\ 58\ 29.80 ,+27\ 18\ 23.5&       17.53&  29199 &  33 & \mathrm{W+S}\\
49            & 7\ 58\ 30.83 ,+27\ 13\ 33.5&       17.44&  29486 &  33 & \mathrm{W+S}\\
50            & 7\ 58\ 33.27 ,+27\ 02\ 18.1&       16.41&  29734 &  29 & \mathrm{W+S}\\
\textit{51}*  & 7\ 58\ 33.99 ,+27\ 11\ 51.4&       18.89&  69649 & 261 & \mathrm{S}\\
\textit{52}   & 7\ 58\ 36.35 ,+27\ 18\ 42.0&       17.75&  22657 &   6 & \mathrm{W+S}\\
53            & 7\ 58\ 36.45 ,+27\ 11\ 36.8&       18.04&  29153 &  74 & \mathrm{W}\\
\textit{54}   & 7\ 58\ 39.62 ,+26\ 51\ 47.7&       17.77&  36780 &  29 & \mathrm{S}\\
\textit{55}   & 7\ 58\ 40.38 ,+26\ 56\ 39.3&       16.79&  36893 &  21 & \mathrm{W+S}\\
\textit{56}   & 7\ 58\ 41.26 ,+26\ 58\ 14.6&       17.12&  22338 &  26 & \mathrm{W+S}\\
57            & 7\ 58\ 42.45 ,+27\ 16\ 56.0&       17.45&  28757 &  21 & \mathrm{W+S}\\
\textit{58}   & 7\ 58\ 46.68 ,+27\ 34\ 51.9&       15.90&  22936 &  51 & \mathrm{S}\\
59            & 7\ 58\ 46.99 ,+27\ 05\ 15.6&       17.15&  29621 &  52 & \mathrm{S}\\
\textit{60}   & 7\ 58\ 47.87 ,+26\ 47\ 35.8&       16.96&  22240 &  17 & \mathrm{S}\\
61            & 7\ 58\ 47.96 ,+27\ 17\ 06.0&       16.33&  28705 &  31 & \mathrm{W+S}\\
62            & 7\ 58\ 49.67 ,+27\ 27\ 55.9&       17.42&  29039 &  35 & \mathrm{S}\\
\textit{63}   & 7\ 58\ 52.31 ,+27\ 24\ 01.6&       18.23&  20449 & 120 & \mathrm{W}\\
64            & 7\ 58\ 52.99 ,+27\ 12\ 51.2&       17.62&  29272 &  30 & \mathrm{W+S}\\
65            & 7\ 58\ 54.80 ,+27\ 11\ 28.5&       17.63&  28444 &  39 & \mathrm{S}\\
\textit{66}   & 7\ 58\ 56.31 ,+27\ 33\ 24.7&       16.44&  22937 &  56 & \mathrm{S}\\
\textit{67}   & 7\ 58\ 56.53 ,+27\ 03\ 17.2&       17.39&  13626 &  12 & \mathrm{S}\\
\textit{68}   & 7\ 58\ 58.20 ,+27\ 09\ 37.1&       15.84&  22629 &  31 & \mathrm{W+S}\\
69            & 7\ 58\ 59.33 ,+27\ 12\ 56.8&       16.68&  28846 &  39 & \mathrm{W+S}\\
\textit{70}   & 7\ 59\ 00.21 ,+27\ 35\ 33.9&       16.92&  14101 &  29 & \mathrm{S}\\

                \noalign{\smallskip}			    
            \hline					    
            \noalign{\smallskip}			    
            \hline					    
         \end{array}
     $$ 
         \end{table}

\addtocounter{table}{-1}
\begin{table}[!ht]
          \caption[ ]{Continued.}
     $$ 
           \begin{array}{r c c r r c}

            \hline
            \noalign{\smallskip}
            \hline
            \noalign{\smallskip}

\mathrm{ID} & \mathrm{\alpha},\mathrm{\delta}\,(\mathrm{J}2000)  & r^\prime & \mathrm{v}\,\,\,\,\,\,\, & \mathrm{\Delta}\mathrm{v}& \mathrm{Source}  \\
%  &                      &  &  &\,\,\,\,\,\,\,\mathrm{(\,km}&\mathrm{s^{-1}\,)}\,\,\,&  & & \\
  & & &\mathrm{(\,km}&\mathrm{s^{-1}\,)}& \\
            \hline
            \noalign{\smallskip}

\textit{71}   & 7\ 59\ 00.70 ,+27\ 33\ 52.8&       17.01&  37403 &  50 & \mathrm{S}\\
72            & 7\ 59\ 01.04 ,+27\ 07\ 12.1&       17.30&  29007 &  29 & \mathrm{W+S}\\
\textit{73}   & 7\ 59\ 01.10 ,+27\ 09\ 43.3&       16.52&  25836 &  26 & \mathrm{W}\\
74            & 7\ 59\ 01.28 ,+27\ 06\ 11.8&       17.36&  29190 &  35 & \mathrm{S}\\
\textit{75}   & 7\ 59\ 03.63 ,+26\ 53\ 09.7&       17.05&  22372 &  28 & \mathrm{W+S}\\
76            & 7\ 59\ 04.34 ,+27\ 08\ 55.4&       16.86&  28833 &  40 & \mathrm{S}\\
77            & 7\ 59\ 04.34 ,+27\ 13\ 25.6&       16.83&  29178 &  38 & \mathrm{W+S}\\
\textit{78}   & 7\ 59\ 05.16 ,+27\ 27\ 34.3&       15.32&   6741 &  54 & \mathrm{S}\\
79            & 7\ 59\ 05.81 ,+27\ 06\ 29.8&       16.77&  29277 &  35 & \mathrm{W+S}\\
\textit{80}   & 7\ 59\ 06.22 ,+27\ 15\ 39.4&       16.54&  22300 &  78 & \mathrm{W}\\
\textit{81}   & 7\ 59\ 06.80 ,+26\ 50\ 54.4&       18.37&  99637 &  46 & \mathrm{S}\\
\textit{82}   & 7\ 59\ 07.49 ,+27\ 18\ 12.0&       17.41&  24710 &  15 & \mathrm{W+S}\\
83            & 7\ 59\ 08.52 ,+27\ 11\ 27.1&       17.41&  29123 &  15 & \mathrm{W+S}\\
\textit{84}   & 7\ 59\ 09.00 ,+27\ 33\ 15.6&       17.44&  23184 &  13 & \mathrm{S}\\
\textit{85}   & 7\ 59\ 10.27 ,+26\ 49\ 07.1&       16.08&  20469 &  44 & \mathrm{S}\\
86            & 7\ 59\ 11.37 ,+27\ 10\ 50.7&       17.31&  30233 &  43 & \mathrm{W}\\
87            & 7\ 59\ 12.89 ,+27\ 06\ 16.0&       16.96&  30172 &  53 & \mathrm{W}\\
88            & 7\ 59\ 13.65 ,+27\ 05\ 41.0&       16.14&  29016 &  40 & \mathrm{W}\\
\textit{89}   & 7\ 59\ 13.80 ,+26\ 38\ 22.8&       17.56&   6473 &  28 & \mathrm{S}\\
90            & 7\ 59\ 16.02 ,+27\ 08\ 57.7&       17.60&  29242 &  23 & \mathrm{S}\\
\textbf{91}   & 7\ 59\ 17.10 ,+27\ 09\ 16.1&       15.19&  29808 & 103 & \mathrm{W}\\
\textit{92}   & 7\ 59\ 17.87 ,+27\ 04\ 40.3&       13.72&   6808 &  50 & \mathrm{S}\\
93            & 7\ 59\ 18.74 ,+27\ 07\ 10.2&       17.42&  27598 &  39 & \mathrm{S}\\
94            & 7\ 59\ 20.05 ,+27\ 09\ 50.4&       16.38&  29693 &  36 & \mathrm{W+S}\\
95            & 7\ 59\ 24.05 ,+27\ 08\ 49.5&       16.24&  29901 &  38 & \mathrm{W}\\
\textit{96}   & 7\ 59\ 24.24 ,+26\ 57\ 32.4&       17.95&   6299 &  56 & \mathrm{W}\\
\textit{97}   & 7\ 59\ 24.36 ,+27\ 17\ 55.4&       17.52&  22809 &  14 & \mathrm{W+S}\\
\textbf{98}   & 7\ 59\ 25.00 ,+27\ 07\ 48.9&       15.72&  27901 &  42 & \mathrm{W+S}\\
99            & 7\ 59\ 27.79 ,+27\ 07\ 03.7&       17.07&  28709 &  26 & \mathrm{W+S}\\
\textit{100}  & 7\ 59\ 27.83 ,+27\ 05\ 45.4&       15.97&  20444 &  30 & \mathrm{W+S}\\
\textit{101}  & 7\ 59\ 28.52 ,+26\ 53\ 22.6&       14.24&   7924 &  23 & \mathrm{S}\\
\textit{102}  & 7\ 59\ 29.71 ,+27\ 01\ 35.2&       12.60&   6654 &  59 & \mathrm{S}\\
\textit{103}  & 7\ 59\ 32.18 ,+26\ 38\ 18.0&       15.87&  22442 &  41 & \mathrm{S}\\
\textit{104}  & 7\ 59\ 32.83 ,+26\ 55\ 08.6&       17.98&  45163 &  79 & \mathrm{W}\\
\textit{105}  & 7\ 59\ 33.50 ,+26\ 43\ 50.3&       16.23&  14062 &  16 & \mathrm{S}\\

              \noalign{\smallskip}			    
            \hline					    
            \noalign{\smallskip}			    
            \hline					    
         \end{array}\\
     $$ 
         \end{table}

\addtocounter{table}{-1}
\begin{table}[!ht]
          \caption[ ]{Continued.}
     $$ 
           \begin{array}{r c c r r c}
            \hline
            \noalign{\smallskip}
            \hline
            \noalign{\smallskip}

\mathrm{ID} & \mathrm{\alpha},\mathrm{\delta}\,(\mathrm{J}2000)  & r^\prime & \mathrm{v}\,\,\,\,\,\,\, & \mathrm{\Delta}\mathrm{v}& \mathrm{Source} \\
%  &                    &  &  &\,\,\,\,\,\,\,\mathrm{(\,km}&\mathrm{s^{-1}\,)}\,\,\,&  & & \\
 & & &\mathrm{(\,km}&\mathrm{s^{-1}\,)}& \\

            \hline
            \noalign{\smallskip}

106           & 7\ 59\ 37.79 ,+26\ 57\ 29.6&       17.43&  29709 &  26 & \mathrm{W+S}\\
\textit{107}  & 7\ 59\ 38.39 ,+27\ 33\ 25.7&       17.17&  24596 &  15 & \mathrm{S}\\
\textit{108}  & 7\ 59\ 40.16 ,+27\ 05\ 14.1&       17.22&  14234 &  20 & \mathrm{W+S}\\
\textit{109}  & 7\ 59\ 40.18 ,+27\ 30\ 30.9&       16.50&  23160 &  41 & \mathrm{S}\\
\textit{110}  & 7\ 59\ 40.90 ,+27\ 35\ 45.2&       16.54&  22861 &  43 & \mathrm{S}\\
\textit{111}  & 7\ 59\ 41.72 ,+27\ 00\ 12.6&       15.43&   6642 &  27 & \mathrm{W+S}\\
112           & 7\ 59\ 42.16 ,+27\ 19\ 17.0&       18.36&  27799 & 134 & \mathrm{W}\\
113           & 7\ 59\ 43.49 ,+27\ 10\ 10.9&       17.13&  29631 &  43 & \mathrm{S}\\
\textit{114}  & 7\ 59\ 44.15 ,+27\ 23\ 13.4&       16.94&  47209 &  40 & \mathrm{S}\\
\textit{115}  & 7\ 59\ 46.49 ,+26\ 38\ 19.2&       17.79&   8121 &  17 & \mathrm{S}\\
\textit{116}  & 7\ 59\ 47.17 ,+27\ 31\ 24.3&       17.58&  31213 &  35 & \mathrm{S}\\
\textit{117}  & 7\ 59\ 48.38 ,+27\ 11\ 37.1&       18.30&  87323 &  34 & \mathrm{S}\\
\textit{118}  & 7\ 59\ 49.80 ,+27\ 09\ 37.1&       17.90&  20839 & 105 & \mathrm{W}\\
\textit{119}  & 7\ 59\ 51.25 ,+26\ 50\ 42.3&       17.12&  49160 &  15 & \mathrm{W+S}\\
\textit{120}  & 7\ 59\ 53.83 ,+26\ 47\ 20.5&       15.12&  14223 &  19 & \mathrm{S}\\
121           & 7\ 59\ 54.14 ,+27\ 01\ 23.5&       17.77&  28725 &  34 & \mathrm{S}\\
122           & 7\ 59\ 54.42 ,+27\ 08\ 54.7&       17.59&  30176 &  34 & \mathrm{S}\\
\textit{123}  & 7\ 59\ 55.14 ,+26\ 50\ 35.2&       17.45&  49281 &  60 & \mathrm{W}\\
\textit{124}  & 7\ 59\ 55.22 ,+27\ 07\ 47.8&       17.31&  13978 &  13 & \mathrm{S}\\
125           & 7\ 59\ 57.75 ,+26\ 53\ 14.7&       17.57&  29480 &  38 & \mathrm{S}\\
\textit{126}  & 7\ 59\ 59.13 ,+27\ 09\ 02.5&       17.52&  20722 &  84 & \mathrm{W}\\
127           & 7\ 59\ 59.84 ,+26\ 57\ 39.0&       18.53&  29882 &  60 & \mathrm{W}\\
\textit{128}  & 8\ 00\ 00.02 ,+27\ 33\ 27.0&       16.35&  31067 &  45 & \mathrm{S}\\
\textit{129}  & 8\ 00\ 00.14 ,+27\ 09\ 00.1&       17.80&  20572 &  12 & \mathrm{S}\\
\textit{130}  & 8\ 00\ 01.06 ,+26\ 39\ 57.1&       14.11&   8129 &  46 & \mathrm{S}\\
\textit{131}  & 8\ 00\ 01.92 ,+27\ 06\ 52.4&       15.95&  12793 &  15 & \mathrm{W+S}\\
\textit{132}* & 8\ 00\ 02.92 ,+27\ 17\ 27.4&       19.96& 567201 & 830 & \mathrm{S}\\
\textit{133}* & 8\ 00\ 06.60 ,+26\ 50\ 54.7&       18.95& 700666 & 404 & \mathrm{S}\\
134           & 8\ 00\ 06.98 ,+27\ 09\ 38.7&       17.37&  29521 &  33 & \mathrm{W+S}\\
135           & 8\ 00\ 07.36 ,+27\ 08\ 19.8&       16.60&  29108 &  40 & \mathrm{W+S}\\
\textit{136}  & 8\ 00\ 07.96 ,+26\ 42\ 24.4&       15.95&  14162 &  43 & \mathrm{S}\\
137           & 8\ 00\ 11.64 ,+27\ 14\ 22.6&       16.99&  29012 &  29 & \mathrm{W+S}\\
\textit{138}  & 8\ 00\ 11.79 ,+27\ 32\ 30.0&       16.28&  30944 &  36 & \mathrm{S}\\
\textit{139}  & 8\ 00\ 14.08 ,+27\ 18\ 35.4&       17.57&  14802 &  15 & \mathrm{S}\\
\textit{140}  & 8\ 00\ 14.30 ,+26\ 41\ 52.8&       14.87&   8407 &  16 & \mathrm{S}\\

              \noalign{\smallskip}			    
            \hline					    
            \noalign{\smallskip}			    
            \hline					    
         \end{array}\\
     $$ 
\end{table}

\addtocounter{table}{-1}
\begin{table}[!ht]
          \caption[ ]{Continued.}
     $$ 
           \begin{array}{r c c r r c}

            \hline
            \noalign{\smallskip}
            \hline
            \noalign{\smallskip}

\mathrm{ID} & \mathrm{\alpha},\mathrm{\delta}\,(\mathrm{J}2000)  & r^\prime & \mathrm{v}\,\,\,\,\,\,\, & \mathrm{\Delta}\mathrm{v}& \mathrm{Source} \\
%  &                      &  &  &\,\,\,\,\,\,\,\mathrm{(\,km}&\mathrm{s^{-1}\,)}\,\,\,&  & & \\
 & & &\mathrm{(\,km}&\mathrm{s^{-1}\,)}& \\
            \hline
            \noalign{\smallskip} 

\textit{141}* & 8\ 00\ 14.90 ,+27\ 20\ 36.2&       19.31&1135030 & 131  & \mathrm{S}\\
142           & 8\ 00\ 15.38 ,+27\ 12\ 50.1&       17.78&  29462 &  65  & \mathrm{W}\\
\textit{143}  & 8\ 00\ 19.77 ,+26\ 42\ 05.2&       13.91&   8335 &  26  & \mathrm{S}\\
\textit{144}  & 8\ 00\ 20.77 ,+27\ 26\ 24.9&       17.63&  58497 &  54  & \mathrm{S}\\
145           & 8\ 00\ 21.25 ,+27\ 09\ 09.8&       17.46&  29279 &  28  & \mathrm{W+S}\\
146           & 8\ 00\ 22.22 ,+27\ 07\ 52.5&       16.27&  29125 &  31  & \mathrm{W+S}\\
147           & 8\ 00\ 22.54 ,+27\ 03\ 08.1&       18.22&  28402 &  91  & \mathrm{W}\\
148           & 8\ 00\ 25.93 ,+27\ 06\ 59.1&       17.06&  28862 &  34  & \mathrm{S}\\
\textit{149}  & 8\ 00\ 30.46 ,+27\ 02\ 13.7&       16.58&  23138 &  19  & \mathrm{S}\\
\textit{150}  & 8\ 00\ 31.67 ,+26\ 45\ 11.7&       16.06&  14240 &  16  & \mathrm{S}\\
\textit{151}  & 8\ 00\ 39.10 ,+27\ 27\ 06.5&       17.76&  11190 &  17  & \mathrm{S}\\
152           & 8\ 00\ 40.35 ,+26\ 59\ 15.7&       17.21&  28800 &  36  & \mathrm{S}\\
\textit{153}  & 8\ 00\ 41.73 ,+27\ 28\ 22.3&       16.11&  14135 &  40  & \mathrm{S}\\
154           & 8\ 00\ 42.93 ,+27\ 13\ 47.7&       16.90&  29235 &  20  & \mathrm{S}\\
\textit{155}  & 8\ 00\ 47.14 ,+27\ 09\ 01.0&       17.11&  23134 &  38  & \mathrm{S}\\
\textit{156}  & 8\ 00\ 48.75 ,+26\ 59\ 10.4&       16.06&   6849 &  40  & \mathrm{S}\\
\textit{157}  & 8\ 00\ 53.09 ,+27\ 13\ 37.6&       16.58&  13917 &  44  & \mathrm{S}\\
\textit{158}* & 8\ 00\ 55.32 ,+27\ 16\ 52.7&       18.84& 532494 & 447  & \mathrm{S}\\
\textit{159}  & 8\ 00\ 57.27 ,+27\ 12\ 28.0&       17.08&  49350 &  37  & \mathrm{S}\\
\textit{160}* & 8\ 00\ 57.36 ,+26\ 57\ 24.1&       18.95& 567456 & 306  & \mathrm{S}\\
\textit{161}  & 8\ 01\ 00.32 ,+27\ 10\ 46.8&       17.00&  14287 &  46  & \mathrm{S}\\
\textit{162}  & 8\ 01\ 14.34 ,+27\ 11\ 12.4&       17.14&  14226 &  13  & \mathrm{S}\\
\textit{163}  & 8\ 01\ 18.51 ,+27\ 12\ 45.5&       14.47&  14101 &  47  & \mathrm{S}\\
\textit{164}  & 8\ 01\ 22.04 ,+27\ 12\ 05.6&       16.21&  14511 &  45  & \mathrm{S}\\
165           & 8\ 01\ 28.93 ,+27\ 04\ 32.3&       16.05&  30157 &  41  & \mathrm{S}\\

              \noalign{\smallskip}			    
            \hline					    
            \noalign{\smallskip}			    
            \hline					    
         \end{array}\\
     $$ 
         \end{table}
%
%\end{document}

A610 is a poor Abell cluster (Abell richness class=0; Abell et
al. \cite{abe89}) located in the background of the rich nearby cluster
ZwCl 0752.9+2833 (at z=0.015--0.027 according to NED).  Kowalski et
al. (\cite{kow84}) reported a $2\sigma$ upper limit for the X--ray
luminosity $L_\mathrm{X}$(2--6 keV)$=0.58\times 10^{44} \ h_{75}^{-2}$
erg\ s$^{-1}$ (in their cosmology). 

We identified the well known bright radio source B2 0756+27 (ID~91 in
our catalog, see below) as a double radio galaxy associated with the
optically brightest, giant elliptical of the cluster (Valentijn
\cite{val79}, see also Owen et al. \cite{owe92}). For this galaxy Owen
et al. (\cite{owe95}) measured a redshift of $z=0.0991\pm0.0002$ and
detected H$\alpha$+NII emission lines. This WAT radio galaxy has been
recently studied by Jetha et al. (\cite{jet06}), too.  The presence of
a diffuse radio source at Southeast from the radio galaxy (see
Fig.~\ref{A610VLA}) was suggested by Valentijn (\cite{val79}) and then
confirmed and classified as a relic by Giovannini \& Feretti
(\cite{gio00}).

Table~\ref{catalog610} lists our velocity catalog and
Fig.~\ref{figimage610} shows the finding chart for the 165 galaxies in
the field of A610. In Table~\ref{catalog610}, for each galaxy we
provide: identification number ID (Col.~1), right ascension and
declination, $\alpha$ and $\delta$ (J2000, Col.~2); $r^{\prime}$ SDSS
magnitudes (Col.~3); heliocentric radial velocities, ${\rm
v}=cz_{\sun}$ (Col.~4) with errors, $\Delta {\rm v}$ (Col.~5); source
of velocity data (W: WHT, S: SDSS, W+S: combined WHT and SDSS) in
Col.~6. The typical error on radial velocity as given by the median
value is 38 \kss.

\subsection{Member selection and global properties}

\begin{figure}[!h]
%\centering
\resizebox{\hsize}{!}{\includegraphics{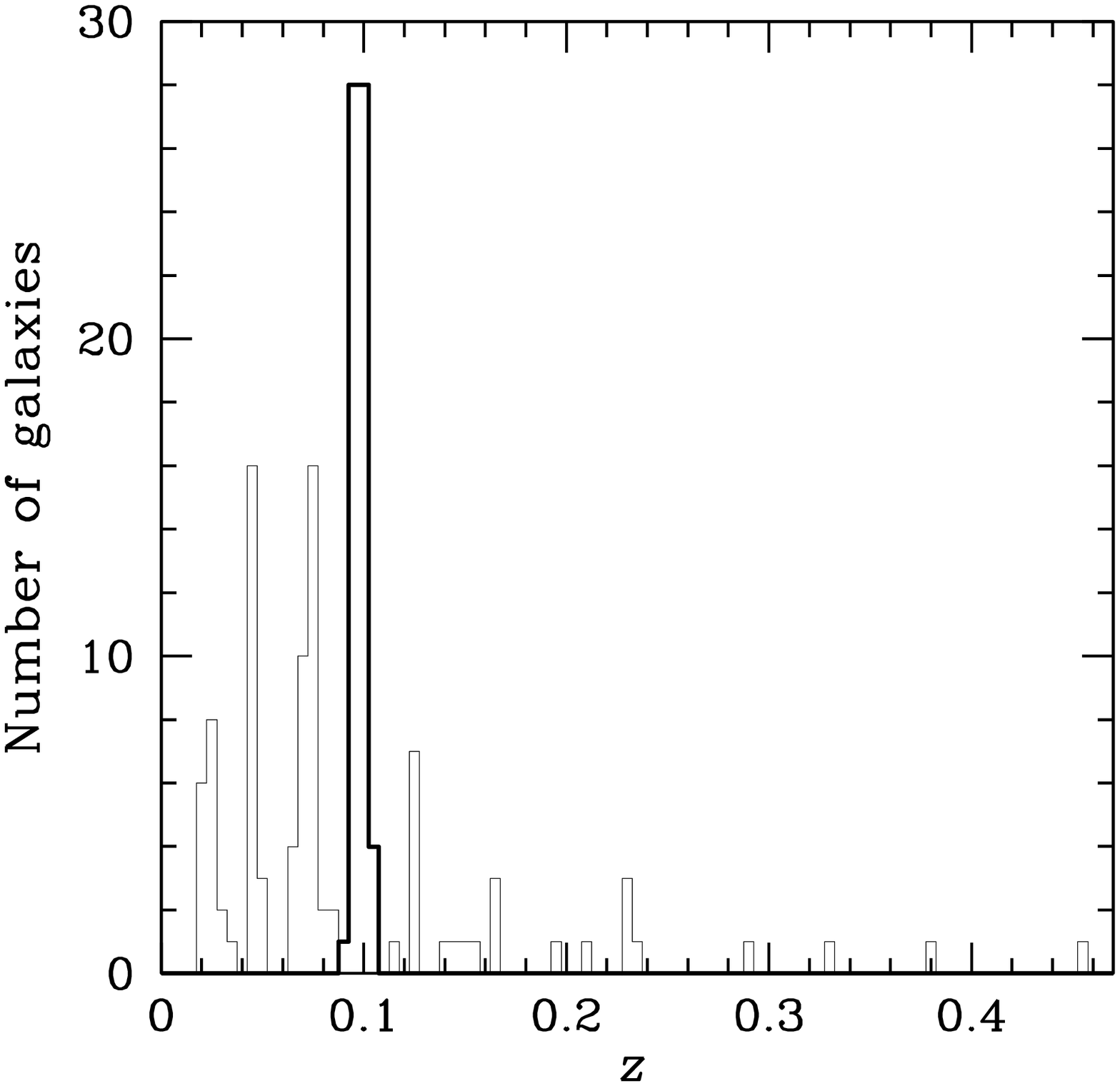}}
\caption
{A610: redshift galaxy distribution. The solid line histogram 
refers to galaxies assigned to the cluster according to
the DEDICA reconstruction method.}
\label{fighisto}
\end{figure}

To select cluster members from the 165 galaxies having redshifts, we
use the adaptive--kernel method (hereafter DEDICA, Pisani \cite{pis93}
and \cite{pis96}; see also Fadda et al. \cite{fad96}; Girardi et
al. \cite{gir96}; Girardi \& Mezzetti \cite{gir01}). We find
significant peaks in the velocity distribution at $>$99\% c.l..  This
procedure detects A610 as a one--peaked structure, populated by 61
galaxies with $27598 \leq v \leq 31213$ \kss (see
Fig.~\ref{fighisto}).  Out of the non--member galaxies, 70 and 34 are
foreground and background galaxies, respectively.

\begin{figure}
%\centering 
\resizebox{\hsize}{!}{\includegraphics{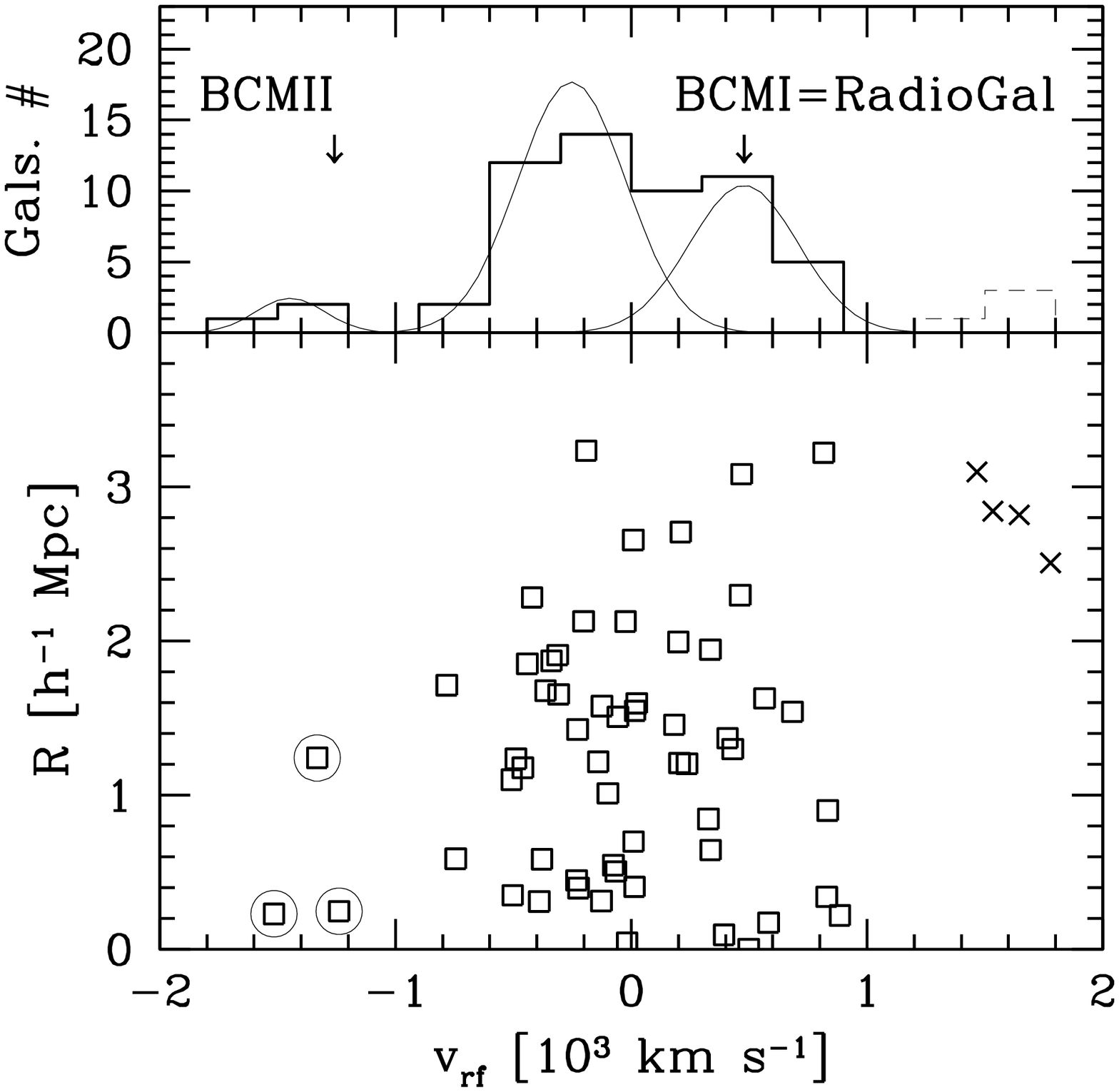}}
\caption
{A610. {\em Bottom panel}: rest--frame velocity vs. projected
clustercentric distance for the 61 galaxies in the main peak
(Fig.~\ref{fighisto}) showing galaxies detected as interlopers by our
``shifting gapper'' procedure (crosses).  Squares indicate galaxies
forming the sample of the 57 member galaxies. The three galaxies with
the lowest velocities are rejected to build Sample2 (see text). {\em
Top panel}: velocity distribution of the 57 cluster members and 4
rejected galaxies (solid and dashed histograms).  Arrows correspond to
the two brightest cluster members. The three Gaussians correspond to
the three KMM groups (see Table~\ref{tabv}).}
\label{figvd}
\end{figure}

All the galaxies assigned to the A610 peak are analyzed in the second
step, which uses the combination of position and velocity information.
We apply the procedure of the ``shifting gapper'' by Fadda et
al. (\cite{fad96}).  This procedure rejects galaxies that are too far
in velocity from the main body of galaxies and within a fixed bin that
shifts along the distance from the cluster center.  The procedure is
iterated until the number of cluster members converges to a stable
value.  Following Fadda et al. (\cite{fad96}) we use a gap of $1000$
\ks in the cluster rest--frame and a bin of 0.6 \hh, or large enough
to include 15 galaxies.  As for the cluster center, we consider the
position of the dominant galaxy
[R.A.=$07^{\mathrm{h}}59^{\mathrm{m}}17\dotsec10$, Dec.=$+27\degree
09\arcmm 16\dotarcs1$ (J2000)] which was identified with the radio
galaxy producing the source B2 0756+27 (see above).  The
shifting--gapper procedure rejects another four galaxies as
non--members (cross symbols in Fig.~\ref{figvd}). Thus, the member
selection procedure leads to a sample of 57 cluster members (see
Table~\ref{catalog610} and Fig.~\ref{figvd}).

\begin{figure}
%\centering 
\resizebox{\hsize}{!}{\includegraphics{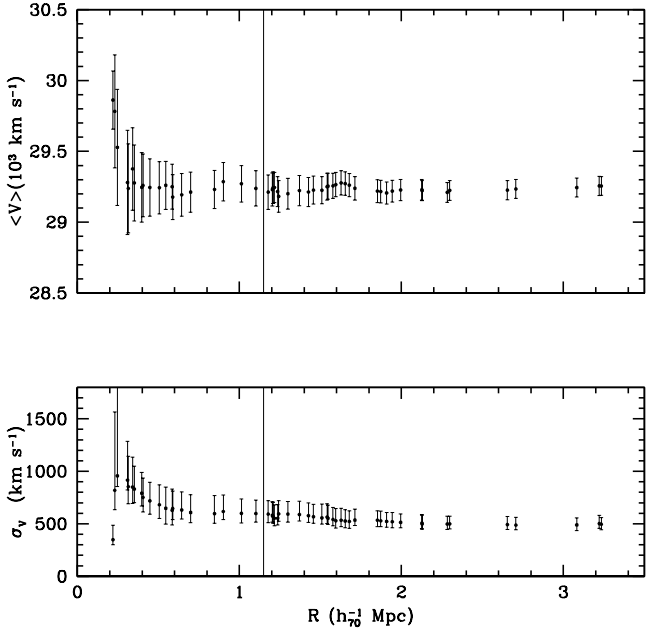}}
\caption
{A610: integral profiles of mean velocity and LOS velocity--dispersion
are shown in {\em top and bottom panels}, respectively.  The mean and
dispersion at a given (projected) radius from the cluster center is
estimated by considering all galaxies within that radius -- the first
value computed on the five galaxies closest to the center. The error
bands at the $68\%$ c.l. are also shown.  The faint vertical lines
give the radius of the virialized region.}
\label{figprof}
\end{figure}

\begin{figure}[!h]
%\centering 
\resizebox{\hsize}{!}{\includegraphics{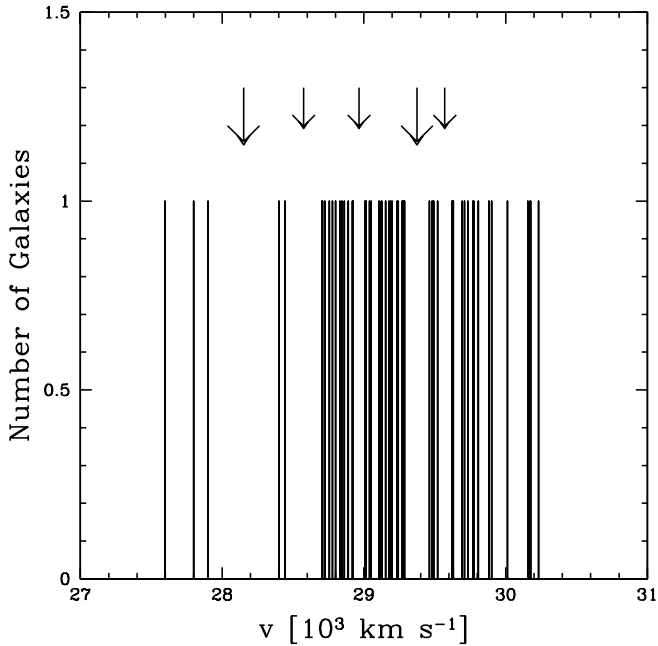}}
\caption
{A610: stripe density plot where the arrows indicate
the positions of the significant gaps. The two most 
important gaps are indicated by the two big arrows.}
\label{figstrip}
\end{figure}

\begin{figure}
%\centering
\resizebox{\hsize}{!}{\includegraphics{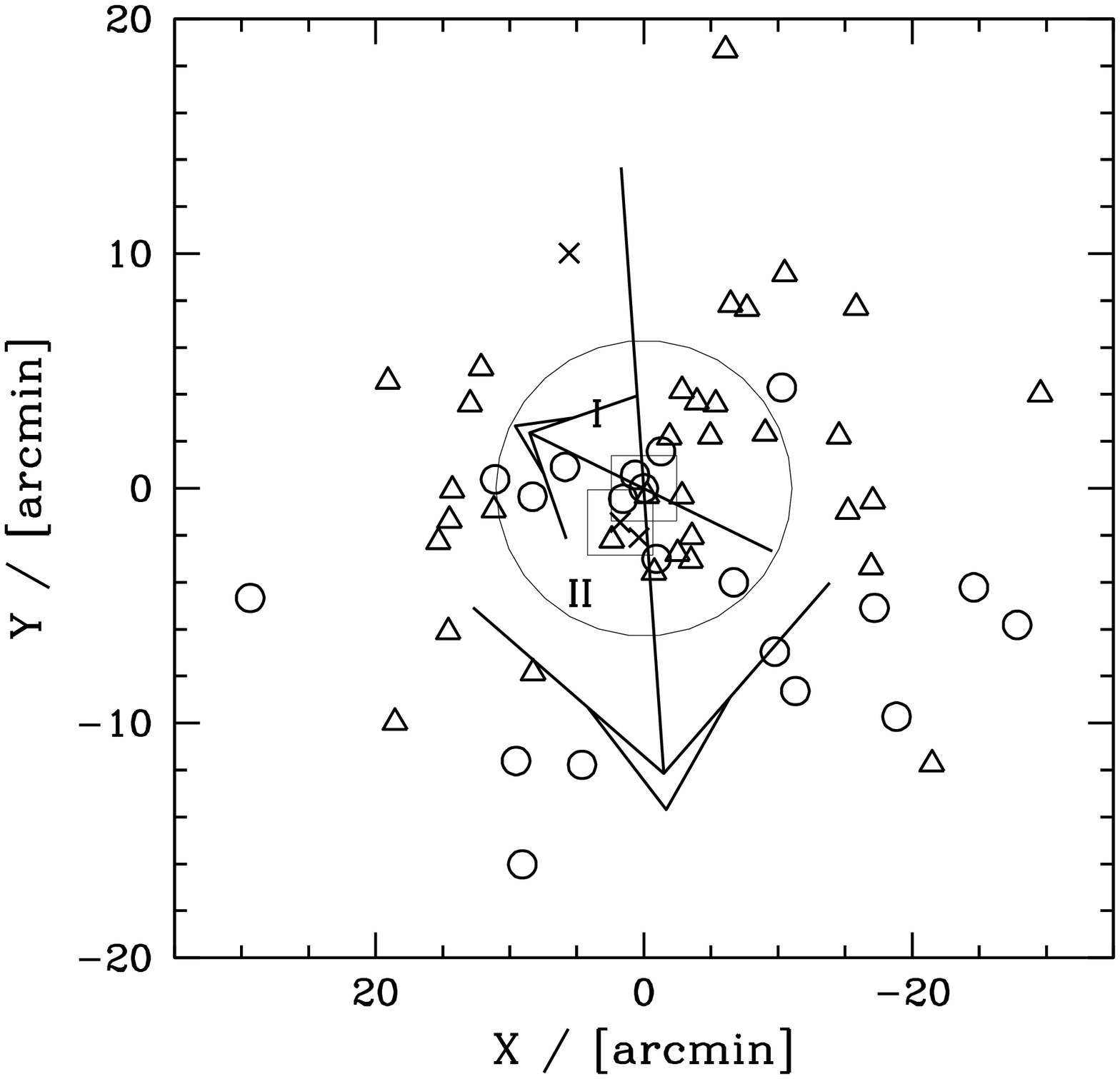}}
\caption
{A610: spatial distribution on the sky of the 57 cluster members.  KMM1,
KMM2 and KMM3 galaxies are denoted by crosses, triangles and circles,
respectively.  The position of the two brightest galaxies (BCMI and
BCMII) are indicated by the faint large squares. The circle indicates
the likely virialized region. Bigger and smaller arrows indicate the
velocity gradients within the whole and the virialized regions,
respectively. The plot is centered on the cluster center defined as
the position of BCMI.}
\label{figkmm}
\end{figure}

By applying the biweight estimator to the cluster members (Beers et
al. \cite{bee90}), we compute a mean cluster redshift of
$\left<z\right>=0.0976\pm$ 0.0002, i.e., $\left<\rm{\rm
v}\right>=29255\pm$66 \kss.  We estimate the LOS velocity dispersion,
$\sigma_{\rm V}$, by using the biweight estimator and applying the
cosmological correction and the standard correction for velocity
errors (Danese et al. \cite{dan80}).  We obtain $\sigma_{\rm
V}=496_{-48}^{+73}$ \kss, where errors are estimated through a
bootstrap technique.

Here we compute the mass of A610 assuming that the system is in
dynamical equilibrium.  Following the prescriptions of Girardi \&
Mezzetti (\cite{gir01}) -- in particular see their Eq.~1 after
introducing the scaling with $H(z)$ -- we assume for the radius of the
quasi--virialized region R$_{\rm vir}=0.17\times \sigma_{\rm V}/H(z) =
1.15$ \h (see also Eq.~ 8 of Carlberg et al. \cite{car97} for
R$_{200}$). Therefore, our spectroscopic catalog samples the whole
virialized region of the cluster.

One can compute the mass using the virial theorem (Limber \& Mathews
\cite{lim60}; see also Girardi et al. \cite{gir98}) under the
assumption that mass follows galaxy distribution: $M=3\pi/2 \cdot
\sigma_{\rm V}^2{\rm R}_{\rm PV}/G$ is the standard virial mass,
R$_{\rm PV}$ a projected radius (equal to two times the harmonic
radius).  Figure~\ref{figprof} shows that the estimate of $\sigma_{\rm
V}$ is robust when computed within a large cluster region, thus we
consider the global value (see Fig.~5 of Girardi et al. \cite{gir06}
and Fadda et al.  \cite{fad96} for other examples).  The value of
R$_{\rm PV}$ depends on the size of the region considered so that the
computed mass increases (but not linearly) with the increasing region
considered.  Considering the 24 galaxies within R$_{\rm vir}$ we
obtain R$_{\rm PV}=0.87\pm0.11$ \h.

We obtain a virial mass $M(<{\rm R}_{\rm
vir}=1.15\,\hhh)=2.3_{-0.6}^{+0.8}$ \mqua.  Notice that in this case
we do not apply the $20\%$ surface pressure term correction (e.g., The
\& White \cite{the86}; Carlberg et al. \cite{car97}; Girardi et
al. \cite{gir98}) since we use the value of the velocity dispersion as
computed within a very large radius.

We also consider an alternative sample of 54 galaxies -- Sample2 --
rejecting the three galaxies with lowest velocity, which are separated
by a gap of $\sim 450$ \ks in the reference frame from the whole
velocity distribution of the cluster (see Fig.~\ref{figvd}).  Since
this is a poor cluster such a value for a gap is quite anomalous (see
also below Sect.~3.2).  The properties of Sample2 are listed in
Table~\ref{tabv}.

We also present the analysis of the 22 galaxies which are the
subsample of Sample2 contained within R$_{\rm vir}$ (hereafter the
``virialized subsample''), to consider galaxies likely belonging to the
true internal, virialized structure.

\subsection{Analysis of the spectroscopic sample}
\label{sec:vel}

We analyze the velocity distribution to look for possible deviations
from Gaussianity that could provide important signatures of complex
dynamics. For the following tests the null hypothesis is that the
velocity distribution is a single Gaussian.

We estimate three shape estimators, i.e., the kurtosis, the skewness,
and the scaled tail index (see, e.g., Beers et al.~\cite{bee91}).  The
value of the skewness (-0.530) shows a marginal evidence that the
velocity distribution differs from a Gaussian at the $90-95\%$
c.l. (see Table~2 of Bird \& Beers~\cite{bir93}).  Moreover, the
W--test (Shapiro \& Wilk \cite{sha65}) marginally rejects the null
hypothesis of a Gaussian parent distribution at the $90\%$ c.l.. The
analysis of Sample2 also shows a marginal sign of non--Gaussianity
having the kurtosis a value of 3.693 (at $90-95\%$ c.l.).

Then we investigate the presence of gaps in the velocity distribution.
A weighted gap in the space of the ordered velocities is defined as
the difference between two contiguous velocities, weighted by the
location of these velocities with respect to the middle of the
data. We obtain values for these gaps relative to their average size,
precisely the midmean of the weighted--gap distribution. We look for
normalized gaps larger than 2.25 since in random draws of a Gaussian
distribution they arise at most in about $3\%$ of the cases,
independently of the sample size (Wainer and Schacht~\cite{wai78}; see
also Beers et al.~\cite{bee91}). We find five significant gaps (see
Fig.~\ref{figstrip}), the main one lying in the body of the
distribution at v$\sim 29500$ \ks and the secondary one separates the
three galaxies with the lowest velocity from the body of the velocity
distribution.  In Table~\ref{tabgap} we list the number of galaxies
and the velocity of the object preceding the gap, the normalized size
(i.e., the "importance'') of the gap itself, and the probability of
finding a normalized gap of this size with the same position in a
normal distribution (as computed with ROSTAT package, Beers et
al. \cite{bee90}). The importance of the main gap is confirmed by the
analysis of Sample2.

We use the results of the gap analysis to determine the first guess
when using the Kaye's mixture model (KMM) to find a possible group
partition of the velocity distribution (as implemented by Ashman et
al. \cite{ash94}). The KMM algorithm fits an user--specified number of
Gaussian distributions to a dataset and assesses the improvement of
that fit over a single Gaussian. In addition, it provides the
maximum--likelihood estimate of the unknown n--mode Gaussians and an
assignment of objects into groups.  KMM is most appropriate in
situations where theoretical and/or empirical arguments indicate that
a Gaussian model is reasonable.  The Gaussian is valid in the case of
cluster velocity distributions, where gravitational interactions drive
the system toward a relaxed configuration with a Gaussian velocity
distribution.  However, one of the major uncertainties of this method
is the optimal choice of the number of groups for the partition.
Using the results of the gap analysis we try to fit two and three
velocity groups on the basis of the two most important gaps.

We find that a two--groups partition is a significantly better
descriptor of the velocity distribution with respect to a single
Gaussian at the $90-92\%$ c.l. (homoscedastic and heteroscedastic
cases) where the first group is given by the three lowest velocity
galaxies. Moreover, we find a significant three--groups partition of
3--34--20 galaxies (at the $96.8\%$ c.l. in the homoscedastic case).
The KMM analysis of Sample2 confirms the presence of the two groups
with 34 and 20 galaxies (at the $\sim 90\%$ c.l.). Table~\ref{tabv}
lists the results for the kinematical analysis of the groups with 3,
34, and 20 galaxies and Fig.~\ref{figvd} shows the corresponding
Gaussians. We also list the virial radii and masses of the
groups KMM2 and KMM3. They are computed assuming the dynamical
equilibrium and adopting an alternative estimate for ${\rm R}_{\rm
PV}$ useful when the centers of the systems are not well defined (see
Barrena et al. \cite{bar07b}, their Sect.~4). However, we point out
that the uncertainty in KMM membership assignments leads to an
artificial truncation of the tails of the velocity distributions, thus
the velocity dispersions of the two groups are underestimated (see
Bird \cite{bir94}). As a consequence, the masses of KMM2 and KMM3
should be considered only approximated lower limits.

\begin{table}
        \caption[]{A610: Results of the kinematical analysis}
         \label{tabv}
                $$
         \begin{array}{l r l l c c}
            \hline
            \noalign{\smallskip}
            \hline
            \noalign{\smallskip}
\mathrm{Sample} & \mathrm{N_g} & \phantom{249}\mathrm{<v>}\phantom{249} & 
\phantom{24}\sigma_{\rm v}^{\mathrm{a}}\phantom{24}&\mathrm{R_{vir}}&\mathrm{Mass(<R_{vir})}\\
& &\phantom{249}\mathrm{km\ s^{-1}}\phantom{249}&\phantom{2}\mathrm{km\ s^{-1}}\phantom{24}&\mathrm{Mpc}&10^{14}\mathrm{M}_{\odot}\\
            \hline
            \noalign{\smallskip}
\mathrm{Whole\ system}         &57 &29255\pm66 &496_{-48}^{+73}&1.15&2.3^{+0.8}_{-0.5}\\
\mathrm{Whole\ system\ Sample2}&54 &29283\pm58 &426_{-39}^{+38}&0.99&1.8^{+0.4}_{-0.4}\\
\mathrm{Virialized\ subsample} &22 &29312\pm106 &480_{-59}^{+71}&1.11&1.8^{+0.6}_{-0.5}\\
\mathrm{KMM1} &3 &27766\pm89 &(118)&-&-\\
\mathrm{KMM2} &34 &29017\pm40 &227_{-21}^{+38}&0.53&0.22_{-0.07}^{+0.09}\\
\mathrm{KMM3} &20 &29767\pm54 &235_{-29}^{+32}&0.54&0.25_{-0.07}^{+0.09}\\
              \noalign{\smallskip}
            \hline
            \noalign{\smallskip}
            \hline
         \end{array}
$$
\begin{list}{}{}  
\item[$^{\mathrm{a}}$] We use the biweigth estimator by
Beers et al. (1990). For the sample with 3 galaxies we indicate the 
standard 
dispersion estimate.
\end{list}
         \end{table}

Notice that the first and the second brightest galaxies of the cluster
(galaxy IDs 91 and 99; hereafter we refer to them as BCMI and BCMII)
are assigned to KMM3 and KMM1 groups.  In particular, BCMI has a
velocity very close to the mean velocity of its host KMM3 system,
while it shows evidence of peculiarity according to the Indicator test
by Gebhardt \& Beers (\cite{geb91}, at the $>95\%$ c.l.) with respect
to the whole system. As for KMM2, it does not host very
luminous galaxies. Indeed, not all galaxy systems are characterized by
particularly bright galaxies. In particular, dominant galaxies are
often absent in poorly evolved structures (see, for instance, the
``irregular'' and ``flat'' clusters according to the definition of
Struble \& Rood (\cite{str82}) and (\cite{str84})).
The existence of a correlation between positions and velocities of
cluster galaxies is a footprint of real substructures.  To investigate
the velocity field of the A610 complex we divide galaxies in a low and
a high velocity samples by using the median cluster velocity and check
the difference between the two distributions of galaxy positions.  The
two distributions are different at the $92.3\%$ c.l.  according to the
2DKS--test. The same analysis on Sample2 leads to a difference at the
$98.4\%$ c.l..  In order to estimate the direction of the velocity
gradient we perform a multiple linear regression fit to the observed
velocities with respect to the galaxy positions in the plane of the
sky (see also den Hartog \& Katgert \cite{den96}; Girardi et
al. \cite{gir96}). We find a position angle on the celestial sphere of
$PA=184_{-19}^{+25}$ degree (measured counter--clockwise from north,
i.e., higher--velocity galaxies lie in the southern region), but
marginally significant (at the $90\%$ c.l.).

Looking for a correlation between positions and velocities, we also
compare the KMM groups two--by--two by applying the 2DKS--test to the
galaxy position of member galaxies.  When comparing KMM2 with KMM3 we
obtain a difference at the $99.88\%$ c.l. and a marginal difference
when comparing KMM1 with KMM2 (at the $90\%$ c.l.). Figure~\ref{figkmm}
shows as the higher velocity KMM3 group lies SW with respect KMM2
group.

\begin{table}
        \caption[]{A610: Results of the weighted--gap analysis}
         \label{tabgap}
                $$
         \begin{array}{l c l c c}
            \hline
            \noalign{\smallskip}
            \hline
            \noalign{\smallskip}
\mathrm{Sample} & \mathrm{N_{gals,prec}} & \mathrm{v_{prec}}& \mathrm{Size}& \mathrm{Prob.of.detect.}\\
                &                  &\mathrm{km\ s^{-1}}&              &
         \\
            \hline
            \noalign{\smallskip}
 
\mathrm{Whole\ sample} & 3  &27901&2.79& 6.0E-3\\
\mathrm{Whole\ sample} & 5  &28444&2.55& 1.4E-2\\
\mathrm{Whole\ sample} & 17 &28924^{\mathrm{a}}&2.32& 3.0E-2\\
\mathrm{Whole\ sample} & 36 &29287^{\mathrm{a}}&3.56& 5.0E-4\\
\mathrm{Whole\ sample} & 5  &29521^{\mathrm{a}}&2.73& 1.4E-2\\
\mathrm{Virialized\ sample} & 14  &29277^{\mathrm{b}}&3.19& 2.0E-3\\
              \noalign{\smallskip}
            \hline
            \noalign{\smallskip}
            \hline
         \end{array}
$$
\begin{list}{}{}  
\item[$^{\mathrm{a}}$] Gaps also found in Sample2.
\item[$^{\mathrm{b}}$] This gap corresponds to the most
significant in the whole sample. 
\end{list}
         \end{table}

When limiting our analysis to the ``virialized subsample'' we find
again: a marginal evidence of non--Gaussianity (at the 90-95$\%$
c.l. from the kurtosis of -0.849); the gap at $\sim 29500$ \kss; a
different spatial distribution between low and high velocity
populations (at the $\gtrsim 95\%$ c.l.).  The presence of a velocity
gradient is now significant at the $94\%$ c.l. with a value of
$PA=64_{-42}^{+59}$ and, in fact, the high velocity KMM3 galaxies lie
NE in the central cluster region (see Fig.~\ref{figkmm}).

\subsection{2D galaxy distribution}
\label{sec:2D}

\begin{figure}
%\centering
\includegraphics[width=9cm]{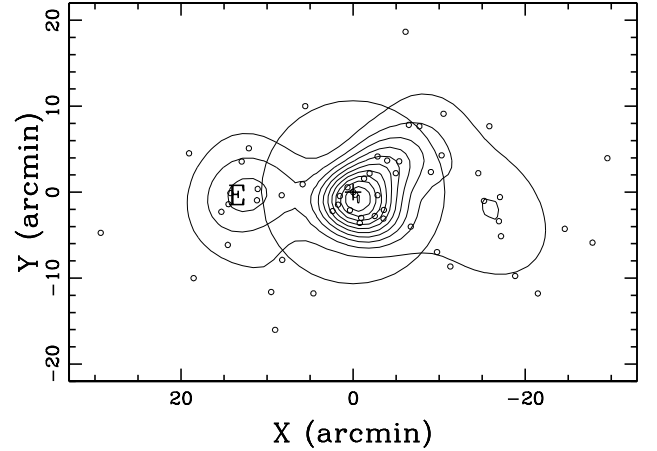}
\caption
{A610: spatial distribution on the sky of spectroscopically confirmed
cluster members and the relative isodensity contour map.  The
brightest cluster member BCMI is indicated by a cross.  The circle
indicates the likely virialized region. The Eastern peak is indicated,
too. The plot is centered on the cluster center defined as the
position of BCMI.}
\label{figk2z}
\end{figure}

When applying the DEDICA method to the 2D distribution of the 57
cluster members we find two very significant peaks (at the $>99.9$
c.l., see Fig.~\ref{figk2z}). The position of the highest peak is
close to the location of the brightest cluster member. The secondary
peak lies at $\sim 15$\arcm East ($\sim 1.6$ \hh) with respect to the
BCMI. We compare the mean velocity of galaxies belonging to the main
peak of our 2D analysis with that of galaxies belonging to the eastern
secondary peak.  We find no difference. This explains why the eastern
structure is not detected in the analyses of Sect.~3.2.

\begin{figure}
%\centering
\includegraphics[width=8cm]{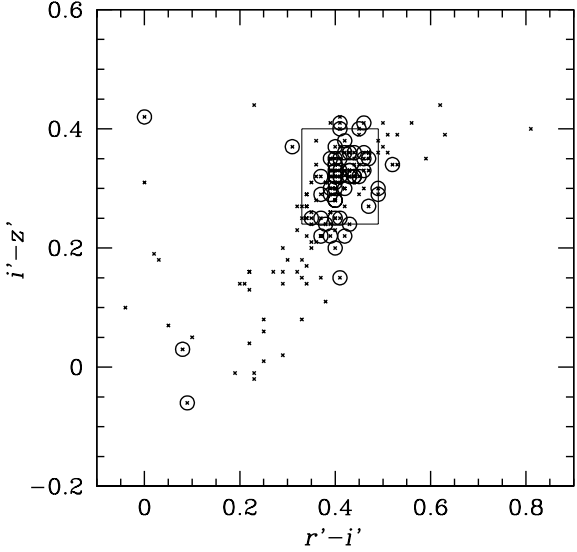}
\caption
{A610: $i'$--$z'$ vs. $r'-i'$ diagram for galaxies with available
spectroscopy is shown by small crosses.  The faint square is centered
on the median value for colors of member galaxies (circles) and
encloses galaxies having colors in the range of 0.08 mag from median
values.}
\label{figcc}
\end{figure}

\begin{figure}
%\centering
\includegraphics[width=8cm]{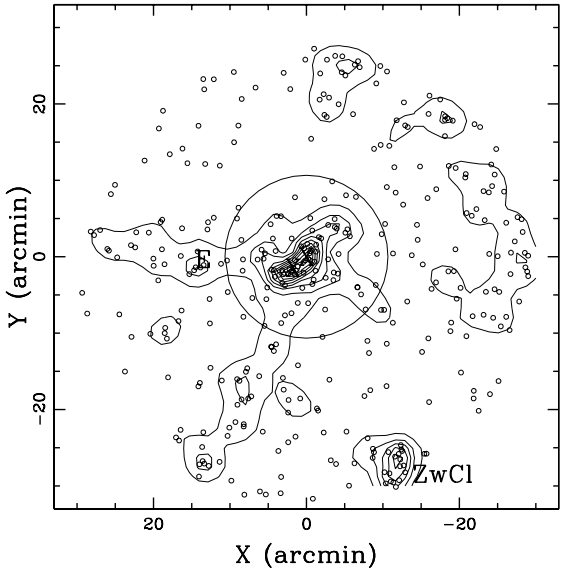}
\caption
{A610: spatial distribution on the sky and relative isodensity contour
map of the 357 likely cluster members (according to the color--color
diagram) with $r'\le 20$, obtained with the DEDICA method. BCMI and
BCMII are indicated by crosses. The circle indicates the likely
virialized region. The Eastern peak and the Zw Cl0755.2+2649 are
indicated, too. }
\label{figk2cc}
\end{figure}

\begin{figure}[!t]
%\centering
\resizebox{\hsize}{!}{\includegraphics{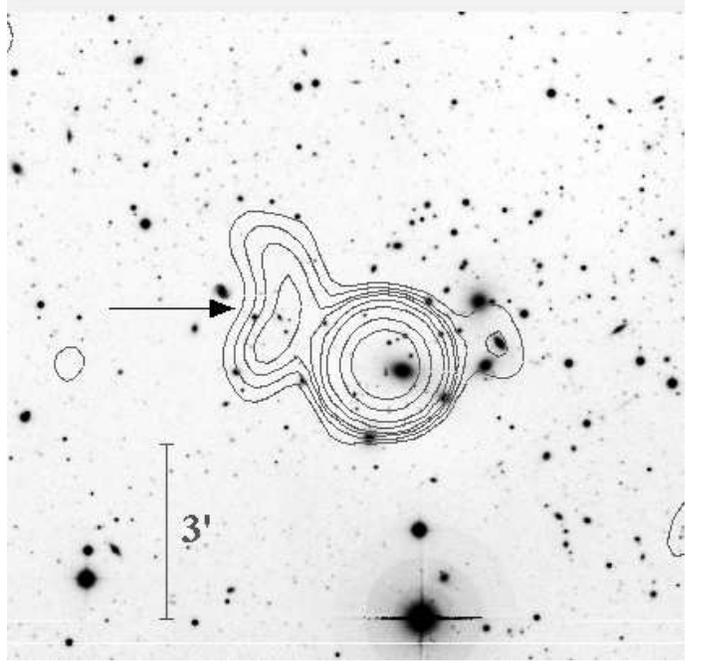}}
\caption
{$R$--band image of A725 (data taken with the WFC camera of the INT)
with, superimposed, the contour levels of a radio image from the
Westerbork Northern Sky Survey (WENSS; see Kempner \& Sarazin
\cite{kem01}). The arrow shows the position of the radio relic. North
is at the top and East to the left.}
\label{A725wenss}
\end{figure}

Our spectroscopic data suffer from magnitude incompleteness.  To
overcome this limit we recover the photometric catalog extracted from
the SDSS DR5.

The color--magnitude relation (hereafter CMR), which indicates the
early--type galaxy locus, is usually applied to select likely cluster
members (see, e.g., A725 in this paper and references therein). When
more than two colors are available, it is more effective to select
galaxies in color-color space.  Here we consider the $i'$-$z'$
vs. $r'$-$i'$ plane where a rectangular box seems suitable for galaxy
selection (e.g., Fig.~12 by Goto et al. \cite{goto02}).  In particular,
we consider likely members galaxies having $r'$ within 0.08 mags from
$r'$--$i'$=0.41 and $i'$--$z'$=0.32, i.e., the median values recovered
from the spectroscopically cluster members (see Fig.~\ref{figcc}).
The value of 0.08 mag is about two times the typical scatter reported
by Goto et al. (\cite{goto02}) for the two CMRs $r'$--$i'$ vs. $r'$
and $i'$--$z'$ vs. $r'$.  To avoid contamination by field galaxies we
do not show results for galaxies fainter than 20 mag (in $r'$--band).

The contour map for the 357 likely cluster members having $r'\le 20$
is shown in Fig.~\ref{figk2cc}.  We also analyze galaxies likely
members having $r'\le 19$ and $19< r'\le 20$ in a separate way.  Our
general results are the following.  We confirm the presence of the
eastern peak, which is generated by bright galaxies.  The main
structure, contained in the virialized region, shows a much more
complex structure having two peaks centered around BCMII and BCMI.

Finally, we notice that the southern peak detected at the southern
border of the field is Zwicky Cluster 0755.2+2649 (recognized by NED
as the RXCJ 0758.3 at z=0.2315, B\"ohringer et al.~\cite{boh00}).

%
% ########################## A725 #################################
%

\section{Abell 725}

%\documentclass{aa}
%\usepackage{graphicx}
%%%new commands
%\def\lesssim{\mathrel{\hbox{\rlap{\hbox{\lower4pt\hbox{$\sim$}}}\hbox{$<$}}}}
%\def\gtrsim{\mathrel{\hbox{\rlap{\hbox{\lower4pt\hbox{$\sim$}}}\hbox{$>$}}}}
%\newcommand{\mincir}{\raise -2.truept\hbox{\rlap{\hbox{$\sim$}}\raise5.truept
%\hbox{$<$}\ }}
%\newcommand{\magcir}{\raise -2.truept\hbox{\rlap{\hbox{$\sim$}}\raise5.truept
%\hbox{$>$}\ }}
%\newcommand{\siml}{\raise -2.truept\hbox{\rlap{\hbox{$\sim$}}\raise5.truept
%\hbox{$<$}\ }}
%\newcommand{\simg}{\raise -2.truept\hbox{\rlap{\hbox{$\sim$}}\raise5.truept
%\hbox{$>$}\ }}
%\newcommand{\be}{\begin{equation}}
%\newcommand{\ee}{\end{equation}}
%\newcommand{\ba}{\begin{eqnarray}}
%\newcommand{\ea}{\end{eqnarray}}
%\newcommand {\h} {$h^{-1}$ Mpc $ \;$}
%\newcommand {\kpc} {$h^{-1}$ kpc}
%\newcommand {\hh} {$h^{-1}$ Mpc}
%\newcommand {\ks} {km~s$^{-1} \;$}
%\newcommand {\kss} {km~s$^{-1}$}
%\newcommand {\mpc} {$Mpc \;$}
%\newcommand {\msun} {$h^{-1} \  M_{\odot} \;$}
%\newcommand {\m} {$M_{\odot} \;$}
%\newcommand {\ml} {$h \, M_{\odot}/L_{\odot} \;$}
%\newcommand {\mll} {$h \, M_{\odot}/L_{\odot}$}
%\newcommand{\vel}{\,{\rm km\,s^{-1}}}
%%%
%\begin{document}
%%
%\addtocounter{table}{-2}
\begin{table}[!ht]
        \caption[]{Velocity catalog of 51 spectroscopically measured
galaxies in the field of A725. In Col.~1, IDs in italics indicate non--cluster
galaxies. ID~24 (in boldface) is BCMI (see text).}
         \label{catalog725}
              $$ 
        % \begin{array}{p{0.5\linewidth}l}
           \begin{array}{r c c c r r}
            \hline
            \noalign{\smallskip}
            \hline
            \noalign{\smallskip}

\mathrm{ID} & \mathrm{\alpha},\mathrm{\delta}\,(\mathrm{J}2000)  & R & B & \mathrm{v}\,\,\,\,\,\, & \mathrm{\Delta}\mathrm{v} \\
  &                      &  &  &\,\,\,\,\,\,\,\mathrm{(\,km}&\mathrm{s^{-1}\,)}\,\,\,\\
            \hline
            \noalign{\smallskip}  

\textit{1}      & 08\ 58\ 55.32 ,+62\ 44\ 53.4&        -  &           -   &  8079&  68 \\
2               & 08\ 58\ 58.09 ,+62\ 36\ 23.5&      16.18&         18.35 & 26850&  39 \\
3               & 08\ 59\ 06.88 ,+62\ 34\ 20.7&      17.73&         19.82 & 27692&  63 \\
4               & 08\ 59\ 17.09 ,+62\ 33\ 21.3&      16.43&         18.07 & 27339&  52 \\
5               & 08\ 59\ 32.75 ,+62\ 20\ 45.9&      16.88&         18.17 & 27389&  68 \\
6               & 08\ 59\ 36.97 ,+62\ 20\ 46.3&        -  &           -   & 27473&  82 \\
\textit{7}      & 08\ 59\ 41.35 ,+62\ 39\ 16.4&      16.53&         18.50 & 49663&  95 \\
8               & 08\ 59\ 47.36 ,+62\ 37\ 58.6&      17.68&         19.14 & 28261&  70 \\
9               & 08\ 59\ 55.76 ,+62\ 35\ 55.6&      17.45&         19.27 & 28561&  90 \\
10              & 09\ 00\ 22.25 ,+62\ 43\ 29.2&      16.61&         18.61 & 26269&  47 \\
\textit{11}     & 09\ 00\ 22.42 ,+62\ 19\ 07.5&        -  &           -   & 13495&  49 \\
12              & 09\ 00\ 25.98 ,+62\ 37\ 45.8&      16.52&         18.69 & 27461&  49 \\
13              & 09\ 00\ 32.55 ,+62\ 42\ 00.8&      16.81&         18.90 & 27293&  60 \\
14              & 09\ 00\ 35.62 ,+62\ 43\ 59.4&      16.83&         18.51 & 27457&  79 \\
\textit{15}     & 09\ 00\ 37.96 ,+62\ 50\ 33.7&      17.37&         19.18 & 50005&  37 \\
16              & 09\ 00\ 38.09 ,+62\ 38\ 05.7&      16.67&         18.66 & 27620& 105 \\
17              & 09\ 00\ 42.36 ,+62\ 34\ 49.2&      16.86&         19.04 & 27414& 117 \\
18              & 09\ 00\ 48.50 ,+62\ 35\ 55.0&      16.79&         18.29 & 28471&  53 \\
19              & 09\ 00\ 55.68 ,+62\ 37\ 49.5&      15.57&         17.64 & 29360&  48 \\
\textit{20}     & 09\ 00\ 55.72 ,+62\ 30\ 06.6&      18.25&         19.84 & 53737&  52 \\
21              & 09\ 00\ 57.04 ,+62\ 32\ 57.8&      15.91&         18.12 & 27235&  24 \\
22              & 09\ 00\ 58.85 ,+62\ 38\ 31.9&      16.09&         16.87 & 27739&  49 \\
23              & 09\ 01\ 03.65 ,+62\ 36\ 52.9&      15.62&         17.67 & 28023&  45 \\
\textbf{24}     & 09\ 01\ 09.99 ,+62\ 37\ 20.0&      15.12&         16.41 & 27104&  48 \\
25              & 09\ 01\ 11.08 ,+62\ 39\ 27.9&      16.61&         18.70 & 28324&  42 \\
26              & 09\ 01\ 14.79 ,+62\ 36\ 10.6&      15.44&         17.59 & 27449&  22 \\
27              & 09\ 01\ 34.70 ,+62\ 37\ 16.1&      16.99&         19.14 & 27616& 103 \\
28              & 09\ 01\ 34.95 ,+62\ 33\ 32.7&      17.56&         18.63 & 25973&  23 \\
\textit{29}     & 09\ 01\ 35.40 ,+62\ 33\ 21.1&      17.75&         19.12 & 30626& 147 \\
30              & 09\ 01\ 36.99 ,+62\ 38\ 38.3&      15.79&         17.80 & 27716&  43 \\
31              & 09\ 01\ 40.93 ,+62\ 37\ 59.9&      17.01&         19.05 & 27942&  83 \\
32              & 09\ 01\ 49.97 ,+62\ 42\ 33.6&      16.37&         18.37 & 27018&  60 \\
33              & 09\ 01\ 52.04 ,+62\ 34\ 42.9&      18.15&         19.20 & 27537& 155 \\
34              & 09\ 01\ 52.15 ,+62\ 34\ 12.3&      16.08&         18.25 & 26873&  75 \\
\textit{35}     & 09\ 02\ 06.50 ,+62\ 54\ 40.7&        -  &           -   & 90943& 101 \\
36              & 09\ 02\ 07.78 ,+62\ 40\ 40.7&      16.27&         18.24 & 26846&  84 \\
\textit{37}     & 09\ 02\ 09.60 ,+62\ 19\ 30.4&        -  &           -   & 11134& 122 \\
38              & 09\ 02\ 09.98 ,+62\ 45\ 01.1&      17.70&         19.65 & 27530&  75 \\
39              & 09\ 02\ 15.21 ,+62\ 40\ 16.4&      14.88&         17.16 & 27108&  48 \\
\textit{40}     & 09\ 02\ 16.28 ,+62\ 32\ 25.7&      17.66&         19.65 & 53311&  57 \\
41              & 09\ 02\ 19.33 ,+62\ 38\ 58.1&      17.40&         19.18 & 27659&  68 \\
42              & 09\ 02\ 19.51 ,+62\ 48\ 14.1&      17.66&         19.16 & 28592&  95 \\
\textit{43}     & 09\ 02\ 19.58 ,+62\ 22\ 16.3&      17.46&         19.74 & 48789&  91 \\
\textit{44}     & 09\ 02\ 20.67 ,+62\ 45\ 44.8&      17.72&         19.58 & 52418&  68 \\
45              & 09\ 02\ 29.76 ,+62\ 28\ 31.3&      16.77&         18.75 & 26878& 130 \\
\textit{46}     & 09\ 02\ 57.75 ,+62\ 22\ 00.1&        -  &           -   & 62471&  71 \\
47              & 09\ 02\ 58.11 ,+62\ 48\ 45.0&        -  &           -   & 27632&  79 \\
\textit{48}     & 09\ 02\ 58.30 ,+62\ 38\ 56.7&        -  &           -   & 22394&  47 \\
\textit{49}     & 09\ 03\ 01.38 ,+62\ 39\ 31.4&        -  &           -   & 22240& 103 \\
50              & 09\ 03\ 06.77 ,+62\ 49\ 39.6&        -  &           -   & 27267&  73 \\
\textit{51}     & 09\ 03\ 13.73 ,+62\ 38\ 32.5&        -  &           -   & 50197&  83 \\

               \noalign{\smallskip}			    
            \hline					    
            \noalign{\smallskip}			    
            \hline					    
         \end{array}
     $$ 
         \end{table}

%\end{document}

A725 is a poor Abell cluster (Abell richness class=0). It is
characterized by an irregular/clumpy galaxy distribution and a low
X--ray luminosity $L_\mathrm{X}$(0.1--2.4 keV)$=0.80\times 10^{44} \
h_{50}^{-2}$ erg\ s$^{-1}$ (B\"ohringer et al. \cite{boh00} in their
cosmology).

The brightest radio source in the cluster (see Fig.~\ref{A725wenss})
is associated with the bright elliptical in the cluster center at
$z=0.0900\pm0.0002$ (Owen et al. \cite{owe93}; \cite{owe95}). The
relic is seen as an arc of diffuse emission to the northeast of this
source (Kempner \& Sarazin \cite{kem01}).  Kempner \& Sarazin also
noticed that the X--ray gas as seen in the ROSAT All--Sky Survey is
slightly elongated along the axis connecting the relic and the cluster
center.

\begin{figure*}[!t]
%\centering
\resizebox{\hsize}{!}{\includegraphics{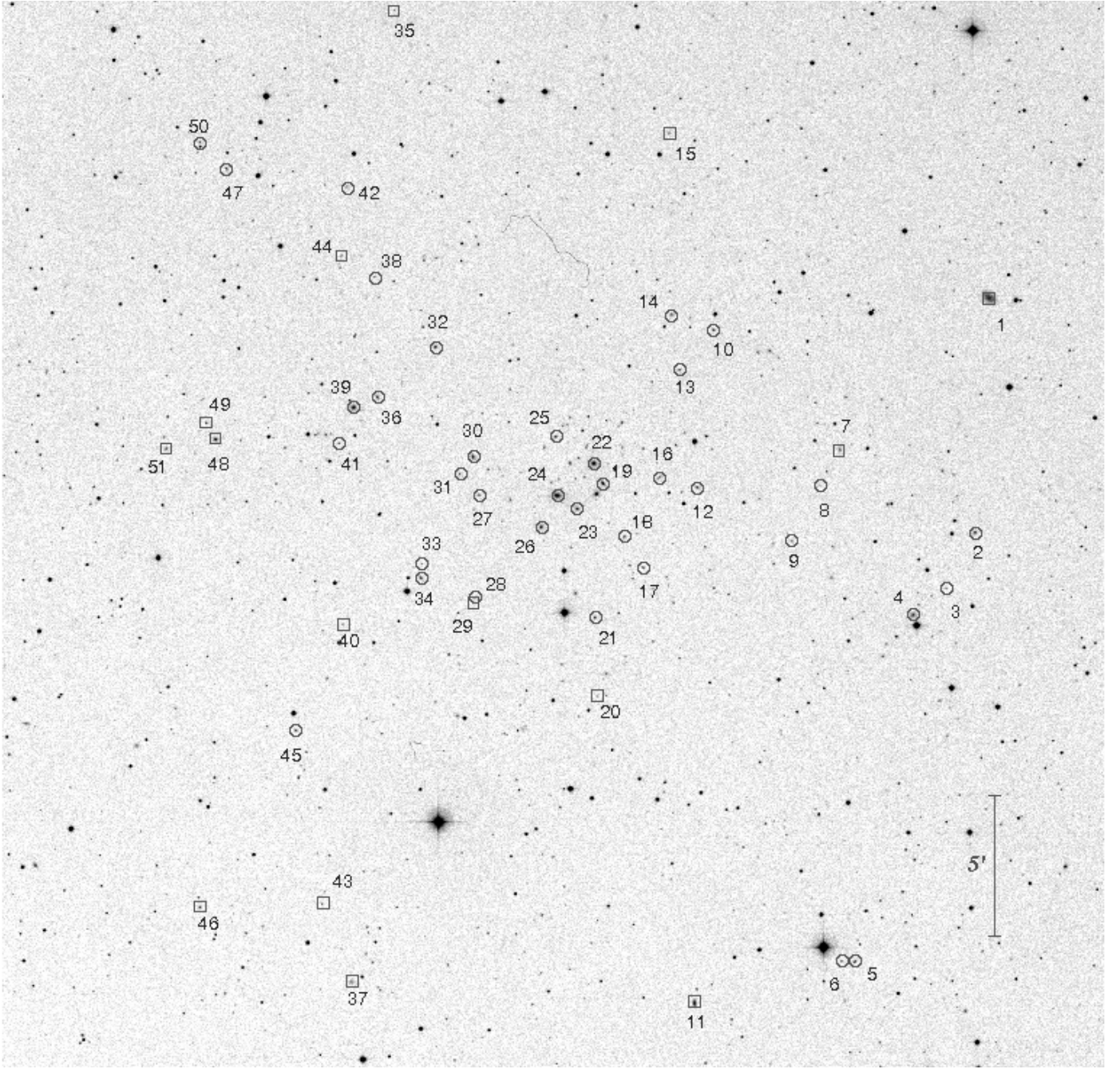}}
\caption{DSS2 $R$--band image of A725 (North at the top and East to the
left). Galaxies with successful velocity measurements are labeled as
in Table~\ref{catalog725}. Circles and boxes indicate cluster members and
non--member galaxies, respectively.}
\label{figimage725}
\end{figure*}

Table~\ref{catalog725} lists our velocity catalog and
Fig.~\ref{figimage725} shows the finding chart for the 51 galaxies in
the field of A725. In Table~\ref{catalog725}, for each galaxy, we
provide: identification number ID (Col.~1), right ascension and
declination, $\alpha$ and $\delta$ (J2000, Col.~2); $R$ and $B$
Johnson magnitudes (Cols.~3 and 4); heliocentric radial velocities,
${\rm v}=cz_{\sun}$ (Col.~5) with errors, $\Delta {\rm v}$
(Col.~6). The typical error on the radial velocity as given by the
median value is 68 \kss.

\subsection{Member selection and analysis}

\begin{figure}
%\centering
\resizebox{\hsize}{!}{\includegraphics{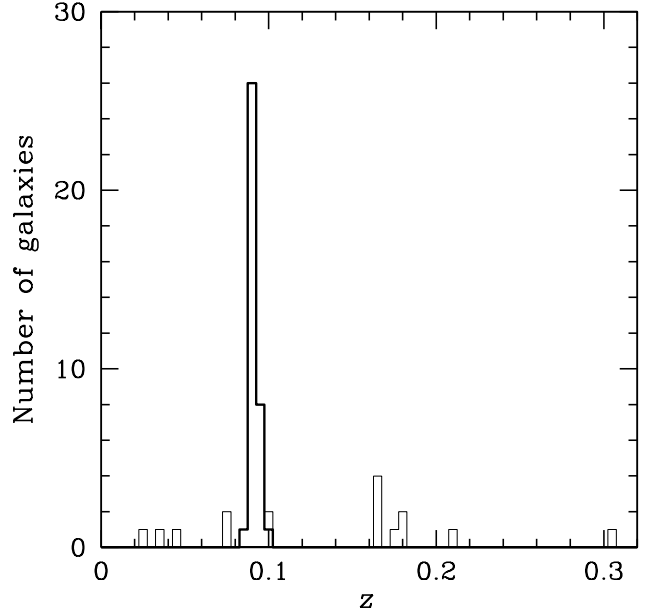}}
\caption
{A725: redshift galaxy distribution. The solid line histogram 
refers to galaxies assigned to the cluster according to
the DEDICA reconstruction method.}
\label{fighisto725}
\end{figure}

Out of 51 galaxies in the sample, the DEDICA method detects A725 as a
one--peaked structure, populated by 36 galaxies with $25973 \leq v \leq 29360$
\kss (see Fig.~\ref{fighisto725}).  Out of the 15 non--member galaxies,
5 and 10 are foreground and background galaxies, respectively.

\begin{figure}
%\centering 
\resizebox{\hsize}{!}{\includegraphics{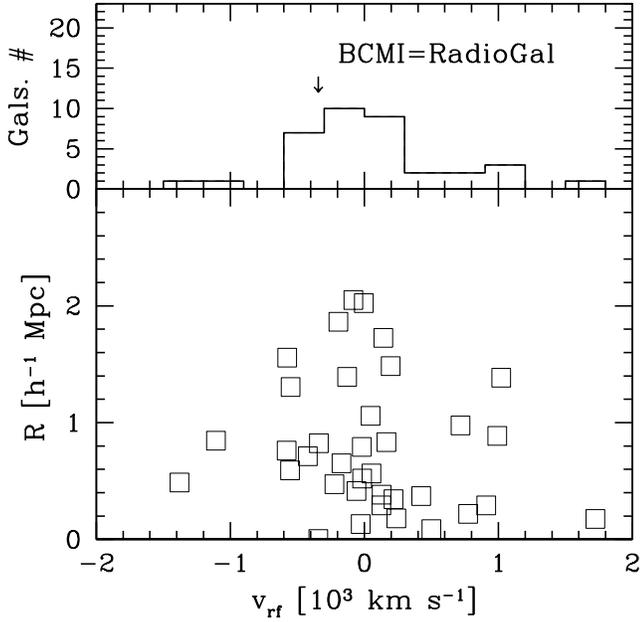}}
\caption
{A725: {\em Bottom panel}: rest--frame velocity vs. projected clustercentric
distance for the 36 member galaxies. {\em Top panel}: velocity
distribution of 36 cluster members. The arrow corresponds to the
brightest cluster member.}
\label{figvd725}
\end{figure}

As for the cluster center, we consider the position of the brightest
cluster member (BCMI; ID~24 in the velocity catalog), which was
identified with the radio galaxy producing the brightest radio source
(see above). No galaxy is rejected by the ``shifting gapper'' leading
to a sample of 36 cluster members (see Table~\ref{catalog725} and
Fig.~\ref{figvd725}).

By applying the biweight estimators to cluster members (Beers et
al. \cite{bee90}), we compute a mean cluster redshift of
$\left<z\right>=0.0917\pm$ 0.0003, i.e., $\left<\rm{\rm
v}\right>=27478\pm$90 \ks and $\sigma_{\rm V}=534_{-97}^{+132}$ \kss.
Assuming that the system is in dynamical equilibrium we compute for
the radius of the virialized region R$_{\rm vir}= 1.24$ \h and
$M(<{\rm R}_{\rm vir}=1.24\,\hhh)=3.2_{-1.2}^{+1.6}$ \mqua.

The values of the kurtosis (3.998) and the scaled tail index (1.504)
indicate that the velocity distribution is lighter--tailed than a
Gaussian at the $90-95\%$ and the $95-99\%$ c.l.s (see Table~2 of Bird
\& Beers~\cite{bir93}), respectively. The BCMI shows evidence of
peculiarity, according to the Indicator test by Gebhardt \& Beers
(\cite{geb91}, at the $>95\%$ c.l.), with respect to the whole system.
We do not find any other trace of substructure.

In this cluster, we find evidence of luminosity segregation.  For the
33 galaxies having available magnitude, the magnitude $R$ correlates
with the clustercentric distance at the $99.5\%$ c.l., according to the
Spearman correlation coefficient.

\subsection{2D galaxy distribution}
\label{sec:2D}

By applying the DEDICA method to the 2D distribution of A725 cluster
members we find that the cluster is elongated along the EW direction.

We recover our photometric catalog selecting likely members on the
basis of the $B$--$R$ vs. $R$ relation (see, e.g., Barrena et
al. \cite{bar07b}).  To determine the relation we fix the slope
according to L\'opez--Cruz et al. (\cite{lop04}, see their Fig.~3) and
apply the two--sigma--clipping fitting procedure to the cluster
members obtaining $B$--$R=2.971-0.054\times R$.  In our photometric
catalog we consider galaxies (objects with SExtractor stellar index
$\le 0.9$) lying within 0.25 mag of the relation.  To avoid
contamination by field galaxies we do not show results for galaxies
fainter than 20 (in the $R$--band).

\begin{figure}
%\centering
\includegraphics[width=8cm]{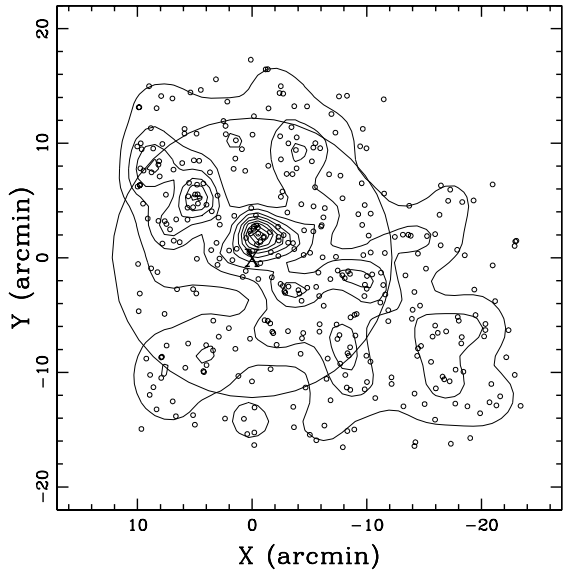}
\caption
{A725: spatial distribution on the sky and relative isodensity contour
map of the 360 likely cluster members (according to the two colors
selection) with $R\le 20$, obtained with the DEDICA method.  The cross
indicates the BCMI. The circle indicates the likely virialized
region.}
\label{figk2cc725}
\end{figure}

The contour map for the 360 likely cluster members having $R\le 20$ is
elongated along the E--W direction in the central region and the
position of the density peak lies about 2\arcm from the BCMI, assumed
as cluster center (see Fig.~\ref{figk2cc725}). Moreover, the cluster
shows an important secondary peak towards NE and the cluster itself
appear elongated along the NE--SW direction (see the isodensity
contours at the virial radius distance).  Similar features are
obtained analyzing other magnitudes cuts ($R\le 19$ or $R\le 18$).

%
% ######################## A796 ######################################
%

\section{Abell 796}

A796 is an Abell cluster of Abell richness class=1. It is
characterized by a clumpy galaxy distribution and a low X--ray
luminosity $L_\mathrm{X}$(0.1--2.4 keV)$=1.38\times 10^{44} \
h_{50}^{-2}$ erg\ s$^{-1}$ (Kempner \& Sarazin \cite{kem01} in their
cosmology).  

%\documentclass{aa}
%\usepackage{graphicx}
%%%new commands
%\def\lesssim{\mathrel{\hbox{\rlap{\hbox{\lower4pt\hbox{$\sim$}}}\hbox{$<$}}}}
%\def\gtrsim{\mathrel{\hbox{\rlap{\hbox{\lower4pt\hbox{$\sim$}}}\hbox{$>$}}}}
%\newcommand{\mincir}{\raise -2.truept\hbox{\rlap{\hbox{$\sim$}}\raise5.truept
%\hbox{$<$}\ }}
%\newcommand{\magcir}{\raise -2.truept\hbox{\rlap{\hbox{$\sim$}}\raise5.truept
%\hbox{$>$}\ }}
%\newcommand{\siml}{\raise -2.truept\hbox{\rlap{\hbox{$\sim$}}\raise5.truept
%\hbox{$<$}\ }}
%\newcommand{\simg}{\raise -2.truept\hbox{\rlap{\hbox{$\sim$}}\raise5.truept
%\hbox{$>$}\ }}
%\newcommand{\be}{\begin{equation}}
%\newcommand{\ee}{\end{equation}}
%\newcommand{\ba}{\begin{eqnarray}}
%\newcommand{\ea}{\end{eqnarray}}
%\newcommand {\h} {$h^{-1}$ Mpc $ \;$}
%\newcommand {\kpc} {$h^{-1}$ kpc}
%\newcommand {\hh} {$h^{-1}$ Mpc}
%\newcommand {\ks} {km~s$^{-1} \;$}
%\newcommand {\kss} {km~s$^{-1}$}
%\newcommand {\mpc} {$Mpc \;$}
%\newcommand {\msun} {$h^{-1} \  M_{\odot} \;$}
%\newcommand {\m} {$M_{\odot} \;$}
%\newcommand {\ml} {$h \, M_{\odot}/L_{\odot} \;$}
%\newcommand {\mll} {$h \, M_{\odot}/L_{\odot}$}
%\newcommand{\vel}{\,{\rm km\,s^{-1}}}
%%%
%\begin{document}
%%
%\addtocounter{table}{-2}
\begin{table}[!ht]
        \caption[]{Velocity catalog of 99 spectroscopically measured
galaxies in the field of A796. In Col.~1, IDs in italics indicate
non--cluster galaxies. IDs~54 and 56 (in boldface) are the brightest
cluster members. Asterisks in Col.~1 highlight QSOs in the SDSS
catalog.}
         \label{catalog796}
              $$ 
        % \begin{array}{p{0.5\linewidth}l}
           \begin{array}{r c c r r c}
            \hline
            \noalign{\smallskip}
            \hline
            \noalign{\smallskip}

\mathrm{ID} & \mathrm{\alpha},\mathrm{\delta}\,(\mathrm{J}2000)  & r^\prime & \mathrm{v}\,\,\,\,\,\,\, & \mathrm{\Delta}\mathrm{v}& \mathrm{Source} \\
  & 09^{\mathrm{h}}      , +60^{\mathrm{o}}    &  &\,\,\,\,\,\,\,\mathrm{(\,km}&\mathrm{s^{-1}\,)}\,\,\,& \\
            \hline
            \noalign{\smallskip}  

%a796.orig                              r'  
\textit{1}*       & 24\ 03.66 , 22\ 43.1 &    18.39&289995&   375 & \mathrm{S}\\       
\textit{2}*       & 24\ 24.87 , 14\ 59.2 &    18.61&644626&   405 & \mathrm{S}\\       
\textit{3}        & 24\ 44.13 , 13\ 53.3 &    17.66& 68395&    54 & \mathrm{S}\\       
\textit{4}        & 24\ 45.38 , 15\ 55.2 &    17.52& 59869&    54 & \mathrm{S}\\       
\textit{5}        & 25\ 07.51 , 04\ 52.1 &    19.27&134817&    84 & \mathrm{S}\\       
\textit{6}*       & 25\ 13.33 , 10\ 39.2 &    17.68&254929&   498 & \mathrm{S}\\       
\textit{7}        & 25\ 26.14 , 31\ 41.4 &    17.49& 77745&    57 & \mathrm{S}\\       
\textit{8}        & 25\ 36.67 , 22\ 35.7 &    17.38& 79529&    48 & \mathrm{S}\\       
9                 & 25\ 39.80 , 07\ 29.9 &    17.23& 47241&    51 & \mathrm{S}\\       
10                & 25\ 50.79 , 17\ 13.2 &    19.38& 47187&    78 & \mathrm{W}\\       
\textit{11}*      & 25\ 59.93 , 26\ 48.5 &    18.17&191738&   492 & \mathrm{S}\\       
12                & 26\ 03.83 , 36\ 02.6 &    17.19& 46654&    45 & \mathrm{S}\\       
\textit{13}*      & 26\ 12.49 , 21\ 02.7 &    19.27&553642&   525 & \mathrm{S}\\       
\textit{14}*      & 26\ 14.75 , 25\ 58.4 &    19.39&670723&   462 & \mathrm{S}\\       
\textit{15}       & 26\ 19.73 , 30\ 15.7 &    18.12& 77784&    45 & \mathrm{S}\\       
\textit{16}       & 26\ 20.66 , 33\ 01.5 &    16.91& 33780&    38 & \mathrm{W+S}\\       
\textit{17}       & 26\ 23.36 , 42\ 20.1 &    17.32& 30615&    18 & \mathrm{S}\\       
18                & 26\ 25.94 , 05\ 09.6 &    17.63& 46216&    30 & \mathrm{S}\\       
\textit{19}       & 26\ 30.12 , 48\ 30.3 &    17.00& 43218&    42 & \mathrm{S}\\       
\textit{20}       & 26\ 30.33 , 28\ 52.9 &    17.96& 90115&    45 & \mathrm{S}\\       
\textit{21}       & 26\ 34.66 , 23\ 39.8 &    17.93& 39946&    57 & \mathrm{W}\\       
\textit{22}       & 26\ 37.08 , 23\ 39.4 &    17.73& 39701&    45 & \mathrm{S}\\       
\textit{23}       & 26\ 39.52 , 37\ 01.0 &    16.97& 21478&    70 & \mathrm{W+S}\\       
\textit{24}       & 26\ 44.25 , 06\ 50.7 &    17.69& 75955&    42 & \mathrm{S}\\       
25                & 26\ 46.11 , 03\ 39.7 &    17.21& 47418&    48 & \mathrm{S}\\       
\textit{26}       & 26\ 46.15 , 31\ 28.8 &    18.41& 90825&    42 & \mathrm{S}\\       
27                & 26\ 49.47 , 27\ 24.2 &    18.22& 45262&    56 & \mathrm{W}\\       
28                & 26\ 49.88 , 02\ 17.2 &    17.70& 47499&    45 & \mathrm{S}\\       
\textit{29}       & 26\ 50.12 , 24\ 36.7 &    17.60& 39648&    42 & \mathrm{S}\\       
\textit{30}       & 26\ 51.06 , 12\ 04.3 &    19.00& 10952&   139 & \mathrm{W}\\       
\textit{31}*      & 26\ 53.67 , 22\ 17.4 &    20.17&455885&   468 & \mathrm{S}\\       
32                & 26\ 53.98 , 19\ 27.0 &    17.91& 47066&    69 & \mathrm{W}\\       
\textit{33}       & 26\ 55.38 , 06\ 49.3 &    18.91& 53621&    64 & \mathrm{W}\\       
\textit{34}*      & 26\ 56.95 , 25\ 04.3 &    19.14&886148& 13317 & \mathrm{S}\\       
\textit{35}       & 26\ 59.88 , 48\ 05.4 &    15.39& 30681&    45 & \mathrm{S}\\       
\textit{36}       & 27\ 01.69 , 03\ 37.3 &    17.67& 53360&    45 & \mathrm{S}\\       
 
               \noalign{\smallskip}			    
            \hline					    
            \noalign{\smallskip}			    
            \hline					    
         \end{array}
     $$ 
         \end{table}
\addtocounter{table}{-1}
\begin{table}[!ht]
          \caption[ ]{Continued.}
     $$ 
           \begin{array}{r c c r r c}
            \hline
            \noalign{\smallskip}
            \hline
            \noalign{\smallskip}

\mathrm{ID} & \mathrm{\alpha},\mathrm{\delta}\,(\mathrm{J}2000)  & r^\prime & \mathrm{v}\,\,\,\,\,\,\, & \mathrm{\Delta}\mathrm{v}& \mathrm{Source} \\
  & 09^{\mathrm{h}}      , +60^{\mathrm{o}}    &  &\,\,\,\,\,\,\,\mathrm{(\,km}&\mathrm{s^{-1}\,)}\,\,\,& \\

            \hline
            \noalign{\smallskip}

37                & 27\ 07.68 , 31\ 50.7 &    17.61&  46872 &  75  & \mathrm{W}\\      
\textit{38}       & 27\ 07.80 , 32\ 27.3 &    17.26&  43198 &  37  & \mathrm{W+S}\\      
\textit{39}       & 27\ 14.41 , 38\ 43.9 &    19.02&  54241 &  58  & \mathrm{W}\\      
\textit{40}       & 27\ 14.95 , 22\ 49.6 &    18.48&  30743 &  48  & \mathrm{W}\\      
\textit{41}       & 27\ 16.64 , 35\ 02.6 &    17.57&  43341 &  33  & \mathrm{W+S}\\      
\textit{42}       & 27\ 25.60 , 22\ 16.2 &    16.46&  30522 &  29  & \mathrm{W+S}\\      
\textit{43}       & 27\ 29.10 , 19\ 51.0 &    17.80&  69182 &  31  & \mathrm{W}\\      
44                & 27\ 29.38 , 27\ 07.8 &    17.72&  46717 &  34  & \mathrm{W}\\      
\textit{45}       & 27\ 34.31 , 11\ 07.5 &    17.05&  22480 &  40  & \mathrm{W+S}\\      
46                & 27\ 40.19 , 27\ 47.2 &    17.58&  46767 &  31  & \mathrm{W+S}\\      
47                & 27\ 40.83 , 26\ 50.4 &    18.16&  45407 & 110  & \mathrm{W}\\      
\textit{48}       & 27\ 42.87 , 40\ 24.9 &    18.65&  33263 & 109  & \mathrm{W}\\      
\textit{49}       & 27\ 44.28 , 08\ 39.3 &    18.29&  21315 &  98  & \mathrm{W}\\      
50                & 27\ 47.09 , 31\ 00.9 &    17.87&  46921 &  43  & \mathrm{W}\\      
\textit{51}       & 27\ 52.10 , 34\ 01.7 &    17.73&  39922 &  83  & \mathrm{W}\\      
\textit{52}       & 27\ 53.74 , 24\ 20.3 &    16.52&   1286 &  39  & \mathrm{S}\\      
53                & 27\ 58.61 , 26\ 17.7 &    17.29&  46961 & 122  & \mathrm{W}\\      
\textbf{54}       & 28\ 00.65 , 26\ 26.8 &    16.96&  46038 &  29  & \mathrm{W+S}\\      
55                & 28\ 03.20 , 24\ 15.7 &    18.72&  47229 &  43  & \mathrm{W}\\      
\textbf{56}       & 28\ 10.75 , 28\ 15.0 &    16.94&  46934 &  39  & \mathrm{W+S}\\      
\textit{57}       & 28\ 11.89 , 45\ 19.8 &    16.94&  39600 &  42  & \mathrm{S}\\      
58                & 28\ 12.80 , 22\ 16.7 &    17.62&  47534 &  37  & \mathrm{W+S}\\      
\textit{59}       & 28\ 14.76 , 02\ 07.5 &    17.59&  30048 &  39  & \mathrm{S}\\      
\textit{60}       & 28\ 19.76 , 49\ 47.3 &    16.26&  26987 &  48  & \mathrm{S}\\      
\textit{61}       & 28\ 24.54 , 14\ 19.4 &    16.97&  22797 &  20  & \mathrm{W+S}\\      
62                & 28\ 24.54 , 23\ 34.8 &    17.84&  47290 &  40  & \mathrm{W}\\      
63                & 28\ 27.29 , 27\ 24.1 &    18.77&  46554 &  58  & \mathrm{W}\\      
\textit{64}       & 28\ 33.51 , 18\ 45.6 &    18.79&  35964 &  66  & \mathrm{W}\\      
\textit{65}*      & 28\ 37.98 , 25\ 21.0 &    16.86&  88574 & 477  & \mathrm{S}\\      
66                & 28\ 41.44 , 05\ 16.6 &    18.40&  48770 &  64  & \mathrm{W}\\      
67                & 28\ 44.90 , 25\ 45.8 &    17.40&  46204 &  39  & \mathrm{S}\\      
\textit{68}       & 28\ 50.43 , 48\ 40.9 &    18.30&  88076 &  48  & \mathrm{S}\\      
\textit{69}       & 28\ 55.52 , 51\ 33.6 &    17.56&  73272 &  45  & \mathrm{S}\\      
\textit{70}       & 28\ 57.65 , 52\ 49.4 &    17.09&  35864 &  45  & \mathrm{S}\\      
                \noalign{\smallskip}			    
            \hline					    
            \noalign{\smallskip}			    
            \hline					    
         \end{array}
     $$ 
         \end{table}
\addtocounter{table}{-1}
\begin{table}[!ht]
          \caption[ ]{Continued.}
     $$ 
           \begin{array}{r c c r r c}
            \hline
            \noalign{\smallskip}
            \hline
            \noalign{\smallskip}

\mathrm{ID} & \mathrm{\alpha},\mathrm{\delta}\,(\mathrm{J}2000)  & r^\prime & \mathrm{v}\,\,\,\,\,\,\, & \mathrm{\Delta}\mathrm{v}& \mathrm{Source} \\
  & 09^{\mathrm{h}}      , +60^{\mathrm{o}}    &  &\,\,\,\,\,\,\,\mathrm{(\,km}&\mathrm{s^{-1}\,)}\,\,\,& \\

            \hline
            \noalign{\smallskip}

\textit{71}       & 28\ 59.48 , 20\ 50.4 &    16.96&   22734 &   37  & \mathrm{W+S}\\
72                & 29\ 15.19 , 14\ 56.9 &    19.36&   47819 &   62  & \mathrm{W}\\
\textit{73}*      & 29\ 18.14 , 17\ 06.3 &    19.03&  665093 &  435  & \mathrm{S}\\
\textit{74}       & 29\ 27.43 , 18\ 03.4 &    17.16&   35941 &   35  & \mathrm{W+S}\\
75                & 29\ 29.47 , 02\ 51.2 &    17.38&   48884 &   42  & \mathrm{S}\\
\textit{76}       & 29\ 32.16 , 37\ 09.1 &    17.13&   35963 &   38  & \mathrm{W+S}\\
\textit{77}       & 29\ 35.13 , 33\ 06.2 &    17.30&   44039 &   41  & \mathrm{W+S}\\
78                & 29\ 50.93 , 14\ 54.6 &    16.92&   47715 &   45  & \mathrm{S}\\
\textit{79}       & 29\ 55.21 , 23\ 17.2 &    16.25&   36079 &   23  & \mathrm{W+S}\\
\textit{80}       & 29\ 58.50 , 12\ 04.6 &    18.39&   80522 &   62  & \mathrm{W}\\
\textit{81}       & 29\ 59.36 , 14\ 46.1 &    17.62&   54170 &   39  & \mathrm{W+S}\\
\textit{82}*      & 30\ 04.28 , 39\ 30.0 &    18.43&  233694 &  345  & \mathrm{S}\\
\textit{83}       & 30\ 06.44 , 26\ 53.4 &    16.93&    4086 &    3  & \mathrm{S}\\
\textit{84}       & 30\ 09.12 , 28\ 05.6 &    16.29&    4072 &   21  & \mathrm{W+S}\\
\textit{85}       & 30\ 14.03 , 21\ 04.6 &    17.66&   42978 &   32  & \mathrm{W+S}\\
\textit{86}       & 30\ 26.63 , 26\ 48.8 &    18.72&   11766 &  100  & \mathrm{W}\\
\textit{87}       & 30\ 30.62 , 22\ 05.0 &    18.34&   50023 &   35  & \mathrm{W}\\
88                & 30\ 33.74 , 12\ 17.6 &    17.56&   46159 &   42  & \mathrm{S}\\
\textit{89}       & 30\ 45.28 , 14\ 23.0 &    17.51&   65622 &   51  & \mathrm{S}\\
\textit{90}*      & 30\ 51.94 , 23\ 01.1 &    19.49& 1111330 &  240  & \mathrm{S}\\
\textit{91}       & 30\ 55.40 , 04\ 36.7 &    19.30&  163600 &   69  & \mathrm{S}\\
\textit{92}       & 31\ 05.10 , 16\ 19.5 &    17.14&   42900 &   42  & \mathrm{S}\\
\textit{93}       & 31\ 05.32 , 14\ 01.1 &    17.30&   53867 &   45  & \mathrm{S}\\
\textit{94}       & 31\ 09.48 , 14\ 59.9 &    15.69&    7765 &   27  & \mathrm{S}\\
\textit{95}       & 31\ 11.54 , 37\ 44.5 &    17.53&   33835 &   33  & \mathrm{S}\\
\textit{96}       & 31\ 13.88 , 06\ 29.1 &    19.12&  187991 &  417  & \mathrm{S}\\
\textit{97}       & 31\ 17.95 , 36\ 03.2 &    16.46&   44219 &   45  & \mathrm{S}\\
\textit{98}       & 31\ 25.83 , 16\ 50.4 &    17.50&   43062 &   42  & \mathrm{S}\\
\textit{99}*      & 31\ 30.71 , 11\ 06.3 &    19.00&  436147 &  687  & \mathrm{S}\\
              \noalign{\smallskip}			    
            \hline					    
            \noalign{\smallskip}			    
            \hline					    
         \end{array}\\
     $$ 
         \end{table}

%\end{document}

Kempner \& Sarazin (\cite{kem01}) found a centrally--located diffuse
radio emission which is quite large and has a very low surface
brightness. When converting in our cosmology the X--ray
luminosity and the radio halo power by Kempner \& Sarazin
(\cite{kem01}), we find $L_\mathrm{X}$(0.1--2.4 keV)=0.93$\times
10^{44}\,h_{70}^{-2}$ erg s$^{-1}$ and $P_{\rm 1.4\,GHz}=(5.6 \pm
2.1)\times 10^{23}$ W Hz$^{-1}$, respectively.  These nominal values
stay far from the correlation between radio power and X--ray
luminosity as given by Cassano et al. (\cite{cas06}; see their
Fig.~1). However, the halo is uncertain and the error on the radio
power is so large that deeper imaging is needed to confirm the
presence of the radio halo and to measure its flux with greater
precision. Owen et al. (\cite{owe93}) found two close radio galaxies
in the field of this cluster. The radio source A is here identified
with a background AGN (ID~65 at $z\sim0.295$ in our catalog, see
below), while we have no $z$ for the radio source B.

\begin{figure*}
%\centering
\resizebox{\hsize}{!}{\includegraphics{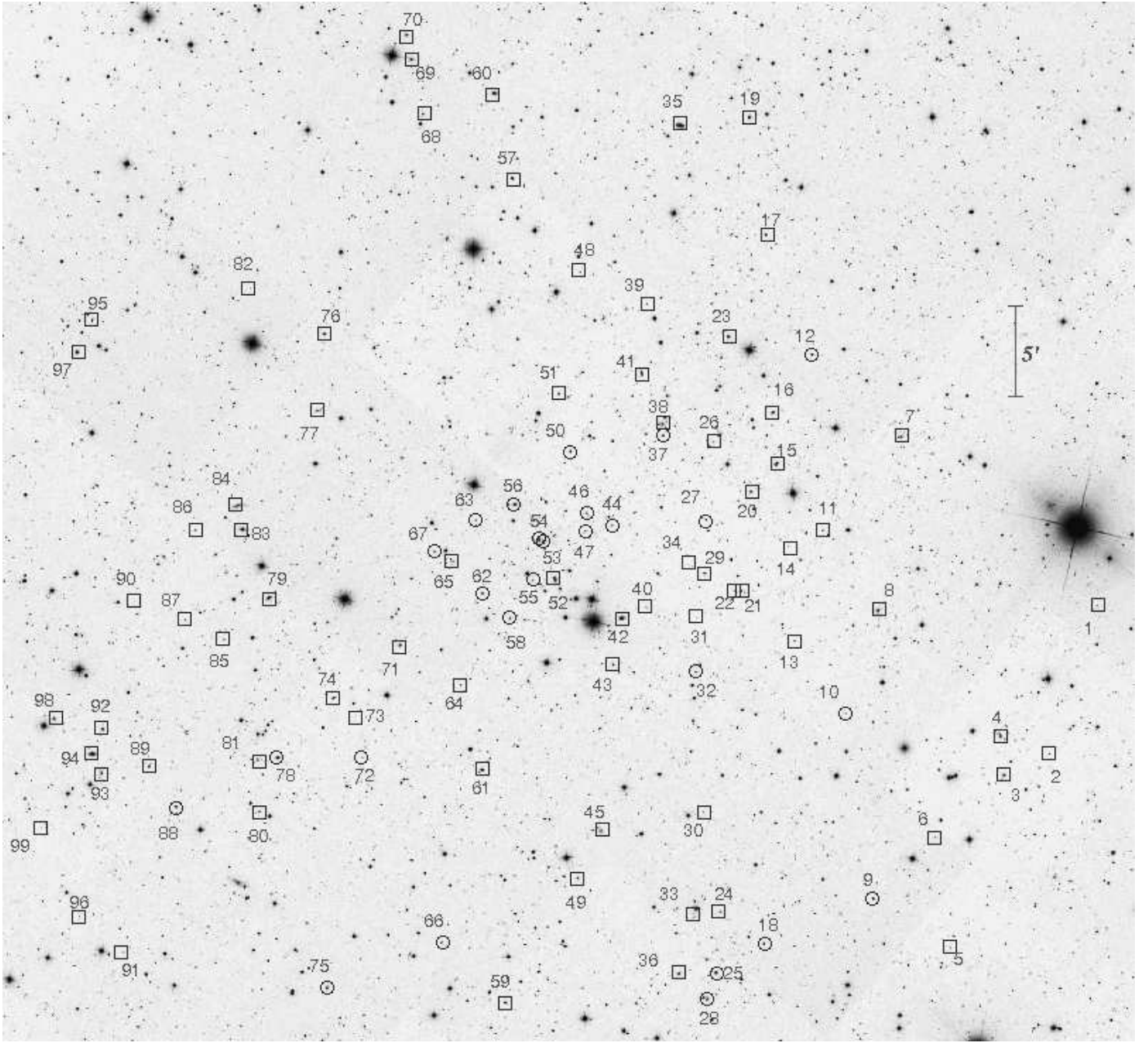}}
\caption
{SDSS image of A796 (North at the top and East to the
left). Galaxies with successful velocity measurements are labeled as
in Table~\ref{catalog796}. Circles and boxes indicate cluster members and
non--member galaxies, respectively.}
\label{figimage796}
\end{figure*}

Table~\ref{catalog796} lists our velocity catalog, and
Fig.~\ref{figimage796} shows the finding chart for the 99 galaxies in
the field of A796. In Table~\ref{catalog796}, for each galaxy we
provide: identification number ID (Col.~1), right ascension and
declination, $\alpha$ and $\delta$ (J2000, Col.~2); $r^{\prime}$ SDSS
magnitudes (Col.~3); heliocentric radial velocities, ${\rm
v}=cz_{\sun}$ (Col.~4) with errors, $\Delta {\rm v}$ (Col.~5); source
of velocity data (W: WHT, S: SDSS, W+S: combined WHT and SDSS) in
Col.~6. The typical error on radial velocity as given by the median
value is 45 \kss.

\subsection{Member selection and analysis}

\begin{figure}
%\centering
\resizebox{\hsize}{!}{\includegraphics{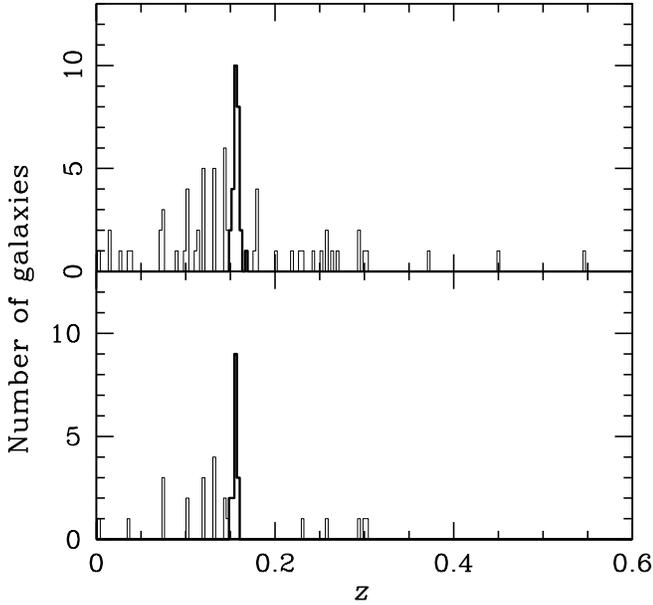}}
\caption
{A796: redshift galaxy distribution of only galaxies with $z<0.6$.
{\em Top panel.} Whole catalog: the solid line histogram refers to
galaxies assigned to the cluster according to the DEDICA
reconstruction method. {\em Bottom panel.} The same for galaxies
within 15\arcm from the cluster center.}
\label{fighisto796}
\end{figure}

Out of 99 galaxies in the sample, the DEDICA method detects A796 as a
one--peaked structure, populated by 27 galaxies with $45242 \leq v
\leq 50023$ \kss (see Fig.~\ref{fighisto796}).  Out of the non--member
galaxies, 38 and 34 are foreground and background galaxies,
respectively.

\begin{figure}[!b]
%\centering 
\resizebox{\hsize}{!}{\includegraphics{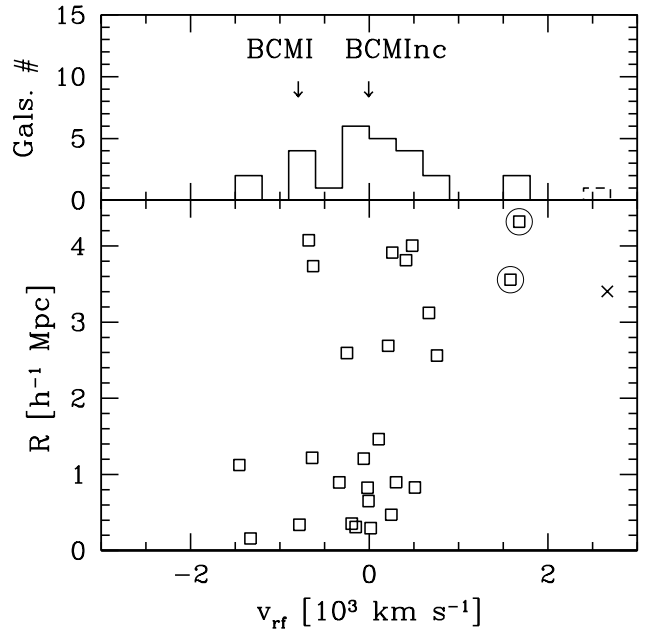}}
\caption
{A796: {\em Bottom panel}: rest--frame velocity vs. projected
clustercentric distance for the 27 galaxies in the main peak
(Fig.~\ref{fighisto}) showing galaxies detected as interloper by our
``shifting gapper'' procedure (crosses).  Squares indicate galaxies
forming the sample of member galaxies. The two galaxies with the
highest velocities are rejected to build Sample2 (see text). {\em Top
panel}: velocity distribution of the 26 cluster members and the rejected
galaxy (solid and dashed histograms).  Arrows correspond to the two
luminous member galaxies.}
\label{figvd796}
\end{figure}

As for the cluster center, we consider the position of the peak in the
2D galaxy distribution as determined by the 2D DEDICA method
[R.A.=$09^{\mathrm{h}}27^{\mathrm{m}}57\dotsec55$, Dec.=$+60\degree
26\arcmm 33\dotarcs7$ (J2000)]. Only one galaxy is then rejected by
the ``shifting gapper'' leading to a sample of 26 cluster members (see
Table~\ref{catalog796} and Fig.~\ref{figvd796}).

Figure~\ref{figvd796} also shows the velocity (with respect to the
cluster mean) of the two luminous cluster galaxies, IDs 56 and
54. They have a comparable $r^\prime$ magnitude, but while ID 54 lies
somewhat closer to the cluster center, ID 56 lies $\lesssim$ 0.4 \h
from the cluster center. Hereafter, we refer to IDs 56 and 54 as BCMInc
and BCMI, respectively.

By applying the biweight estimators to 26 cluster members (Beers et
al. \cite{bee90}), we compute a mean cluster redshift of
$\left<z\right>=0.1566\pm$ 0.0005, i.e., $\left<\rm{\rm
v}\right>=46942\pm$140 \ks and $\sigma_{\rm V}=698_{-159}^{+216}$
\kss. Assuming that the system is in dynamical equilibrium we compute
for the radius of the virialized region R$_{\rm vir}= 1.57$ \h and the
virial mass $M(<{\rm R}_{\rm vir}=1.57\,\hhh)=6.9_{-3.0}^{+4.0}$
\mqua.

\begin{figure}
%\centering 
\resizebox{\hsize}{!}{\includegraphics{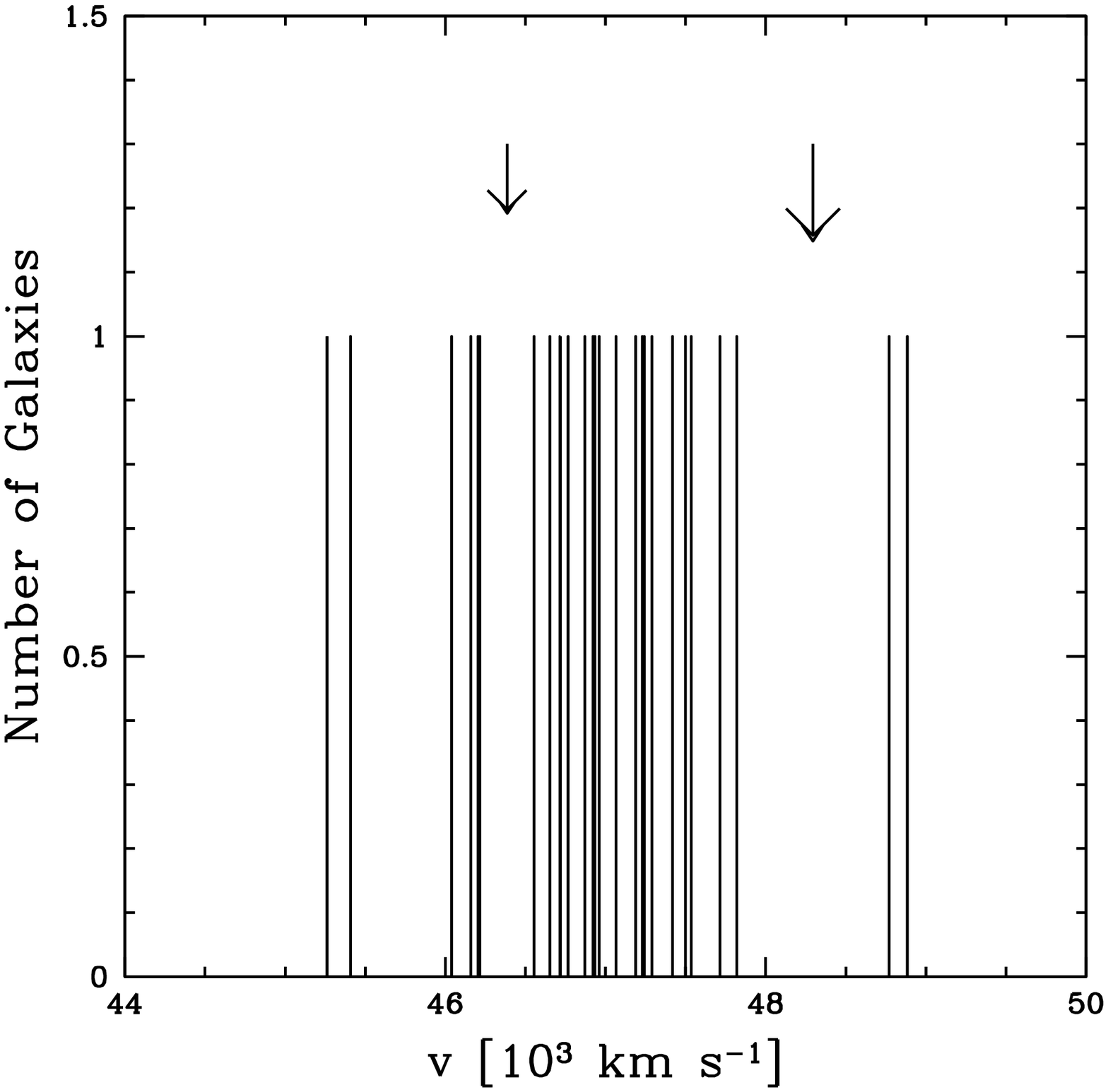}}
\caption
{A796: stripe density plot where the big (small) arrow indicates
the position of the significant gap in the velocity distribution
of the whole system (Sample2).}
\label{figstrip796}
\end{figure}

We also consider an alternative sample of 24 galaxies -- Sample2 --
rejecting the two galaxies with the highest velocity that are
separated by a gap of $\sim 820$ \ks in the reference frame from the
whole velocity distribution of the cluster (see
Fig.~\ref{figvd796}). This gap is detected with a significance level
of (1--3.0E--2) in the velocity distribution (see
Fig.~\ref{figstrip796}). The properties of Sample2 are listed in
Table~\ref{tabv796}.

We also present the analysis of the 14 galaxies which are the
subsample of Sample2 contained within R$_{\rm vir}$ (hereafter the
``virialized subsample'') to consider galaxies likely belonging to the
true internal, virialized structure.

Considering the whole sample of 26 galaxies we find the presence of a
velocity gradient at the $98.9\%$ c.l., with high velocity galaxies
SSE located ($PA=156_{-26}^{+21}$).  Accordingly, the two subsamples
of galaxies having lower or higher velocity than the median velocity
show different distributions in the sky (at the $99.5\%$ according to
the 2DKS test). Moreover, the Dressler--Schectman test finds
substructure at the 96\% c.l.. However, these pieces of evidence of
correlation between velocities and positions are no longer detected
when we analyze Sample2 and the ``virialized sample''. Thus, we
suspect that the correlations found are not really due to cluster
structure, but rather to the large--scale structure environment.  On
the other hand the direction of the velocity gradient coincides with
that shown by the cluster elongation in the 2D analysis (see in the
following). In all our three samples, the BCMI shows evidence of
peculiarity according to the Indicator test by Gebhardt \& Beers
(\cite{geb91}, at the $>95\%$ c.l.) with respect to the whole system.

Considering Sample2, we find a weighted gap which separates the
velocity distribution in a low and a high velocity set (see
Fig.~\ref{figstrip796}). Very interestingly, the two most luminous
galaxies BCMI and BCMInc lie in the low and high velocity sets,
respectively. This is suggestive of a situation similar to that of
A610, i.e., a possible merger between two subclusters.

\subsection{2D galaxy distribution}
\label{sec:2D}

\begin{figure}
%\centering
\includegraphics[width=8cm]{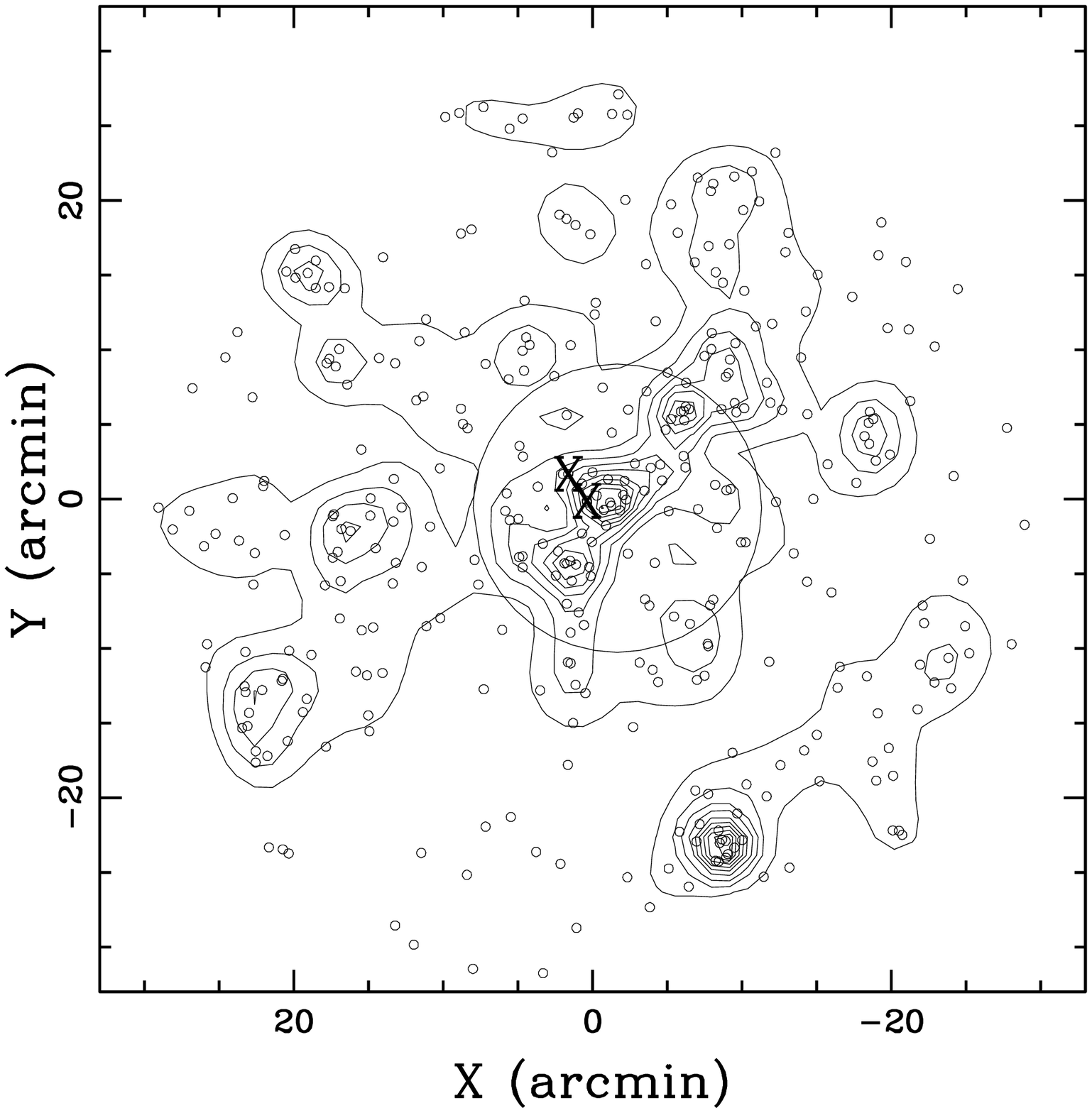}
\caption
{A796: spatial distribution on the sky and relative isodensity contour
map of the 349 likely cluster members (according to the two colors
selection) with $r'\le 20$, obtained with the DEDICA method.  The
circle indicates the likely virialized region. The crosses indicate
the two most luminous galaxies BCMI and BCMInc.}
\label{figk2cc796}
\end{figure}

When applying the DEDICA method to the 2D distribution of A796 galaxy
members we find that the cluster is elongated along the SE--NW
direction. To check this finding we recover the photometric catalog
extracted from the SDSS DR5. Figure~\ref{figk2cc796} shows the result of
the DEDICA method applied to the $r^\prime \leq 20$ likely members
selected on the basis of their colors -- i.e., only galaxies lying
within 0.08 mags from $r'$--$i'$=0.46 and $i'$--$z'$=0.33 colors. We
confirm the cluster elongation finding two minor peaks in the SE and
NW directions. Similar results are found for different magnitude cuts
$r^\prime \leq 19$ and $r^\prime \leq 21$.

\section{Summary \& conclusions}

We present the results of a dynamical analysis of the three poor, low
X--ray luminous clusters A610, A725, and A796 containining diffuse
radio sources (relic, relic, and a possible halo, respectively).  Our
analysis is based on new redshift data for 158 galaxies obtained at
the WHT and additional SDSS DR5 data for A610 and A796.  We also use
new photometric data obtained at the INT for A725.

We select 57, 36 and 26 cluster members for A610, A725 and A796,
respectively. Out of these galaxies 35, 36 and 17 were measured by
the WHT. 

The low values we compute for the global LOS velocity dispersion of
galaxies ($\sigma_{\rm V}=420-700$ \kss) confirm that these clusters
are low--mass clusters. Table~\ref{tabamm} summarizes the main cluster
properties as recovered in this study.  Using the $\sigma_{\rm
V}-T_{\rm X}$ relation by Girardi et al. (\cite{gir98}) the range of
expected X--ray temperature of the intracluster gas is $T_{\rm X}=2-3$
keV.  The study of these three clusters indicates that the phenomenon
of diffuse radio sources is not limited to massive clusters only.

We also discuss the dynamical status and the possible connection with
the diffuse sources of each individual cluster. Regarding the two
clusters hosting a radio relic, A610 and A725, the former is better
sampled and can be analyzed in greater detail. A610 shows a lot of
evidence of substructure: the non--Gaussianity of the velocity
distribution, the correlation between galaxy velocities and positions
and the peculiarity of the BCMI in the velocity space.  A610 seems
formed by two structures separated by $\sim 700$ \ks in the cluster
rest-frame, having comparable $\sigma_{\rm V} \sim 200$ \ks and likely
causing a velocity gradient. Moreover, the BCMI velocity is very close
to the mean velocity of the higher velocity structure.  A third small,
low velocity group hosts the BCMII. The analysis of the 2D galaxy
distribution shows a bimodal distribution in the core -- elongated in
the SE-NW direction and likely associated to BCMI and BCMII groups --
as well as the presence of an eastern group just outside the
virialized region. In this chaotic scenario, it is not easy to unveil
the event responsible for the relic. One possibility is that the relic
be associated to the likely merger between the two structures
corresponding to BCMI and BCMII. In this case it would be somewhat
perpendicular to the axis connecting the two merging structures (see
Fig.~\ref{A610VLA}) in agreement with being originated by shock waves
connected to the ongoing merger (e.g., Ensslin \& Br\"uggen
\cite{ens02}). Another possibility is that the relic be related to the
merger between the two main velocity structures forming the cluster
(see Fig.~\ref{figvd}). This would be more likely from the energetic
point of view since this merger involves structures of 1:1 mass
ratio. In conclusion, A610 represents another case of a radio cluster
formed by two merging subclusters, like, e.g., A115 (Barrena et
al. \cite{bar07b}); A2744 (Boschin et al. \cite{bos06}); and A773
(Barrena et al. \cite{bar07a}), but at much smaller scales (global
$\sigma_{\rm V}\sim 500$ \ks vs. $\sigma_{\rm V}>1000$
\kss). As a further support to the merging scenario, we notice
that the BCMI of A610 is also a bright radio source with a WAT
structure. This is very interesting because WATs are only found in
galaxy clusters and are excellent probes of the gas--dynamical
processes occuring during a cluster merger (e.g., G\'omez et al.
\cite{gom97}).

\begin{table*}
        \caption[]{A796: Results of the kinematical analysis}
         \label{tabv796}
                $$
         \begin{array}{l r l l c c}
            \hline
            \noalign{\smallskip}
            \hline
            \noalign{\smallskip}
\mathrm{Sample} & \mathrm{N_g} & \phantom{249}\mathrm{<v>}\phantom{249} & 
\phantom{24}\sigma_{\rm v}\phantom{24}&\mathrm{R_{vir}}&\mathrm{Mass(<R_{vir})}\\
& &\phantom{249}\mathrm{km\ s^{-1}}\phantom{249} &\phantom{2}\mathrm{km\ s^{-1}}\phantom{24}&\mathrm{Mpc}&10^{14}\mathrm{M}_{\odot}\\
            \hline
            \noalign{\smallskip}
\mathrm{Whole\ system}         &26 &46942\pm140 &698_{-159}^{+216}&1.57&6.9^{+4.0}_{-3.0}\\
\mathrm{Whole\ system\ Sample2}&24 &46921\pm120 &571_{-73}^{+134}&1.29&3.9^{+2.4}_{-2.2}\\
\mathrm{Virialized\ subsample} &14 &46796\pm165 &586_{-148}^{+171}&1.32&4.6^{+2.4}_{-1.6}\\
              \noalign{\smallskip}
            \hline
            \noalign{\smallskip}
            \hline
         \end{array}
$$
         \end{table*}

\begin{table*}
        \caption[]{Properties of galaxy clusters}
         \label{tabamm}
$$
         \begin{array}{l r r r c c c c}
            \hline
            \noalign{\smallskip}
            \hline
            \noalign{\smallskip}
\mathrm{Cluster} & \mathrm{N} & \mathrm{N_{m}}&\mathrm{N_{m,vir}} & \alpha\,\,\,\,\,\,\,\,\,\,\,\,\, \mathrm{\delta}&z&\sigma_{\rm v}\phantom{24}&\mathrm{Mass(<R_{vir})}\\
&&&&\mathrm{J2000}&&\mathrm{km\ s^{-1}}&10^{14}\mathrm{M}_{\odot}\\ 
            \hline
            \noalign{\smallskip}
\mathrm{A610}&165&57&22&075917.10+270916.1&0.098&426\mathrm{-}496&1.8-2.3\\
\mathrm{A725}& 51&36&27&090109.99+623720.0&0.092&\sim534&\sim3.2\\
\mathrm{A796}&99&26&14&092757.55+602633.7&0.157&571\mathrm{-}698&3.9-6.9\\
              \noalign{\smallskip}
            \hline
            \noalign{\smallskip}
            \hline
         \end{array}
$$
         \end{table*}

As for A725, it shows some evidence of non--Gaussianity in the
velocity distribution, the peculiarity of the BCMI in the velocity
space, and an elongated cluster shape in the NE-SW direction.  This
direction is the same one indicated by the radio relic since it is an
arc of diffuse emission to the NE of the radio galaxy associated with
the BCMI (Kempner \& Sarazin \cite{kem01}, see Fig.~\ref{A725wenss}).

A796 possibly hosts a radio halo (to be confirmed) according to WENSS
data (Kempner \& Sarazin \cite{kem01}). It shows the peculiarity of
the BCMI in the velocity space, possible spatial/velocity substructure
and the presence of two minor clumps along the SE--NW direction.

For A725 and A796 the number of cluster members ($\sim 30$) is enough
for the estimate of $\sigma_{\rm V}$, but too small for the analysis of
substructure. Thus, a definitive conclusion will require more data.

\begin{acknowledgements}

We would like to thank Luigina Feretti for many enlightening
discussions and for the VLA radio image of A610 she kindly provided
us. We thank the anonymous referee for useful comments and
suggestions.

This publication is based on observations made on the island of La
Palma with the William Herschel Telescope (WHT) and with the Isaac
Newton Telescope (INT), operated by the Isaac Newton Group (ING), in
the Spanish Observatory of the Roque de Los Muchachos of the Instituto
de Astrofisica de Canarias.

This research has made use of the NASA/IPAC Extragalactic Database
(NED), which is operated by the Jet Propulsion Laboratory, California
Institute of Technology, under contract with the National Aeronautics
and Space Administration.

This research has made use of the galaxy catalog of the Sloan Digital
Sky Survey (SDSS). Funding for the SDSS has been provided by the
Alfred P. Sloan Foundation, the Participating Institutions, the
National Aeronautics and Space Administration, the National Science
Foundation, the U.S. Department of Energy, the Japanese
Monbukagakusho, and the Max Planck Society. The SDSS Web site is
http://www.sdss.org/.

The SDSS is managed by the Astrophysical Research Consortium for the
Participating Institutions. The Participating Institutions are the
American Museum of Natural History, Astrophysical Institute Potsdam,
University of Basel, University of Cambridge, Case Western Reserve
University, University of Chicago, Drexel University, Fermilab, the
Institute for Advanced Study, the Japan Participation Group, Johns
Hopkins University, the Joint Institute for Nuclear Astrophysics, the
Kavli Institute for Particle Astrophysics and Cosmology, the Korean
Scientist Group, the Chinese Academy of Sciences (LAMOST), Los Alamos
National Laboratory, the Max-Planck-Institute for Astronomy (MPIA),
the Max-Planck-Institute for Astrophysics (MPA), New Mexico State
University, Ohio State University, University of Pittsburgh,
University of Portsmouth, Princeton University, the United States
Naval Observatory, and the University of Washington.

This work was partially supported by a grant from the Istituto
Nazionale di Astrofisica (INAF, grant PRIN-INAF2006 CRA ref number
1.06.09.06).
\end{acknowledgements}

\end{document}